\documentclass[12pt]{article}
\usepackage[top=0.75in,left=1in,footskip=0.5in]{geometry}
\usepackage[utf8]{inputenc}
\usepackage{graphicx}
\usepackage{graphics}
\usepackage{geometry}
\usepackage{amsmath}
\usepackage{amsfonts}
\usepackage{amssymb}
\usepackage{latexsym}
\usepackage{setspace}
\usepackage{mathtools}
\usepackage{subcaption}

\usepackage[square,numbers]{natbib}

\usepackage{color}

\usepackage{subfiles}

\usepackage{url}

\usepackage{textcomp}

\usepackage[table]{xcolor}
\usepackage{array}
\usepackage{tabularx}
\usepackage{tfrupee}
\usepackage{newunicodechar}

\usepackage{multirow}
\usepackage[final]{pdfpages}
\usepackage{authblk}
\usepackage{booktabs}
\usepackage{comment}

\begin{document}
	
	\title{Reversing Food Craving Preference Through Multisensory Exposure}
	
	\author[a]{Avishek Chatterjee}
	\author[a]{Satyaki Mazumder}
	
	\author[a]{Koel Das\thanks{Corresponding Author: koel.das@iiserkol.ac.in}}
	\affil[a]{Department of Mathematics and Statistics, Indian Institute of Science Education and Research Kolkata, Nadia, West Bengal-741246, India}

	
	\maketitle  
	
	
	\newpage
	

	\begin{abstract}
		Experiencing food craving is nearly ubiquitous and has several negative pathological impacts, but effective intervention strategies to control or reverse craving remain limited. Food cue-reactivity tasks are often used to study food craving but most paradigms ignore individual food preferences, which could confound the findings. We explored the possibility of reversing food craving preference 
		using psychophysical tasks on human participants considering their individual food preferences in a multisensory food exposure set-up. Participants were grouped into Positive Control (PC), Negative Control (NC), and Neutral Control (NEC) based on their preference for sweet and savory items. Participants reported their momentary craving of the displayed food stimuli through desire scale and bidding scale (willingness to pay) pre and post multisensory exposure. Participants were exposed to food items they either liked or disliked. Our results asserted the effect of the multisensory food exposure showing statistically significant increase in food craving for negative control post-exposure to disliked food items. 
		Using computational model and statistical methods we also show that desire for food does not necessarily translate to willingness to pay every time and instantaneous subjective valuation of food craving is an important parameter for subsequent action.
		Our results further demonstrate the role of parietal N200 and centro-parietal P300 in reversing craving preference. 
	\end{abstract}
	
	
	\section*{Introduction}
	Food preference and consumption constitute a vital aspect of our daily life and have a long-term impact on our health. Food craving, signifying a strong liking for a specific food item, has recently garnered a lot of attention due to its effect in the field of psychology, behavioral economics and medicine. Majority of humans have experienced food craving and the craved items are highly palatable and energy-dense comprising of high fat and/or sugar content \cite{lafay2001gender}. 
	Pathologically, food craving contributes to various negative outcome including obesity, addiction and other disorders like binge eating disorder (BED) and bulimia nervosa \cite{waters2001internal}. Obesity, in particular, deserves special attention due to its unprecedented increase globally.  Most of the public health interventions to sensitize people to healthy food choices in order to control obesity seems to be ineffective. Ultimately choosing to eat healthy food requires self-control over our thought and behavior and effective cognitive strategies can be devised to facilitate the self-control. Previous research has demonstrated that cognitive strategies reduce craving \cite{mischel1975cognitive,wilson2007psychological,giuliani2014neural,boswell2018training} and reduction in craving for non-nutritious food has been shown to be a precondition for healthy eating behavior \cite{jansen2016lab,jansen2010decreased}. 
	
	Developing efficient strategies to control food craving has gained considerable attention from public health perspectives due to increase in craving related disorders in the last decade. There have been a few intervention studies to nudge people towards healthy food by using cognitive \cite{kober2010regulation,svaldi2015effects,meule2013time,sarlo2013cognitive,boswell2018training}, behavioural \cite{liu2014using, schwab2017disgust} and financial strategies \cite{harris2013temporally}. However, due to our poor understanding of the underlying neuro-cognitive mechanism guiding food craving, applicability of efficient intervention strategies remains limited. Our current work attempts to address this issue by exploring behavioral and neural correlates of food craving. 
	
	One of the primary objectives of the current study is to induce a reversal of craving preferences by exposing human observers to food items they dislike using multi-sensory food exposure. Unlike most intervention studies which down-regulate craving \cite{giuliani2013piece,giuliani2014neural}, we wanted to study whether it is possible to up-regulate one's liking for previously disliked food items by use of multi-sensory food exposure. Most food craving studies expose the same food items to participants irrespective of their preferences. Ideally, inducement of craving and its effect would depend on the participant's desire for that particular food item. Exposing participants to a savory when they crave something sweet would not generate the desired effect and will in fact produce erroneous results and confounding interpretations. In this study, participant's food preference was taken into account while exposing them to the food items to explore the effect of positively or negatively desired food items on their behavior. Our study systematically explores the effect of craving using visual food cues before and after food exposure of liked and disliked food items.
	
	Effect of craving is typically transient and transition from craving to spending is best captured momentarily \cite{konova2018computational}. Although craving is known to modulate our valuation system, the effect of craving on consumer behavior remains unexplored in most studies where typically craving is measured by self-reported inconsequential desire rating for the displayed food cues \cite{wolz2017subjective,ledoux2013using}. In the current study, using both desire scale and a bidding scale, we measure the instantaneous desire as well as willingness to pay (WTP) for the food items. WTP is a common measure in the field of behavioral economics to study consumer behavior and helps translate desire for food into subjective valuation \cite{boswell2018training}. Furthermore, to understand the computational mechanism guiding the craving, a detailed statistical model is used which can aid the theoretical understanding of the psychological process of food craving.
	
	Finally, we explore the neural correlates of food craving by using univariate and multivariate analyses of EEG signals recorded during the psychophysical tasks. 
	Event-related potentials (ERPs), with their high temporal resolution, allow identifying the time course of cortical processes of related events with a specific stimulus. 
	Assessing initial sensory or visual attention in response to food cues can elucidate further higher-order attentional processing which may impact on subsequent eating behavior. The ERP components P200/P2, 
	and  posterior N200/N2 
	onset are associated with initial  sensory or visual attention \cite{carbine2018utility}. Greater P2 (or P2 like) amplitude enhancements are associated to food stimuli as compared to neutral stimuli \cite{franken2011electrophysiology} and to chocolate stimuli as compared to non-chocolate stimuli in the group of female chocolate cravers \cite{asmaro2012spatiotemporal}. 
	P300 and late positive potential (LPP) are two commonly examined components that reflect higher-order attentions. P300 
	is commonly associated with automatic attention allocation \cite{zorjan2021changing} and inhibitory control \cite{luijten2014systematic}. Several studies reported higher P300 amplitude to food than non-food stimuli \cite{nijs2008food,lapenta2014transcranial} signifying greater attention allocation of food than non-food stimuli.
	In the addiction literature, less pronounced P300 amplitude in addicted populations as comparison to controls is considered  as a marker for neural deficits in inhibitory control \cite{luijten2014systematic}. 
	LPP on the other hand 
	reflects more extended attention allocation due to the emotional and motivational relevance of pictures which lead to higher amplitudes corresponding to the more arousing stimuli \cite{hajcak2010event,carbine2018utility}.
	In the current study, we compute the known ERP components typically used in craving literature and explore the early and late neural markers that modulate craving of liked and disliked food items.

	\section*{Materials and Methods}
	\label{Method}
	\subsection*{Participants}
	Ninety six adults ($N=96$, 49 female and 47 male,  ages: 18-29, mean: 21.56, std: 2.64)
	participated in two experiments. Exclusion criteria for all participants were: being dieters (medically or self-imposed dietary restrictions), having any history of chronic illness, being younger than 18 years, and being obese (Body Mass Index [BMI] $ \geq $ 30) or underweight (BMI $<$ 18). Also, a participant is excluded from the experiment if his/her time duration of refraining from eating before the experiment is either less than two or more than six hours. Indian Institute of Science Education And Research Kolkata's Ethics Committee approved all  procedures.
	\subsection*{Neuroeconomic Decision-Making Task}
	The pre-questionnaire of participant selection contains likert scales \cite{allen2007likert} for ten food items (five chocolates and five chips items). Each likert scale ranges from the value 1 to 10, where 1 and 10 indicate minimum and maximum desire for the item, respectively. 
	Based on the response, participants were exposed to either one of the food items in the multisensory food exposure of the experiment and thus divided into three groups,  namely Positive Control, PC (exposed a particular food item with specific liking), Neutral Control, NEC (exposed  chocolate or chips, since they are neutral to the both food items) and Negative Control, NC (exposed a particular food item with specific disliking). The details of the above class divisions are mentioned in the Supporting Information.
	The participants were presented with visual stimuli consisting of liked and disliked food items, followed by multisensory exposure to a food item. Participants were again shown visual stimuli following food exposure. On finishing the experiment, the participants answered a post-questionnaire. The study participation consisted of approximately one hour, which was realized in one session. 
	
	
	\begin{figure}[!h]
		\begin{center}
			\includegraphics[width=16cm,height=9cm]{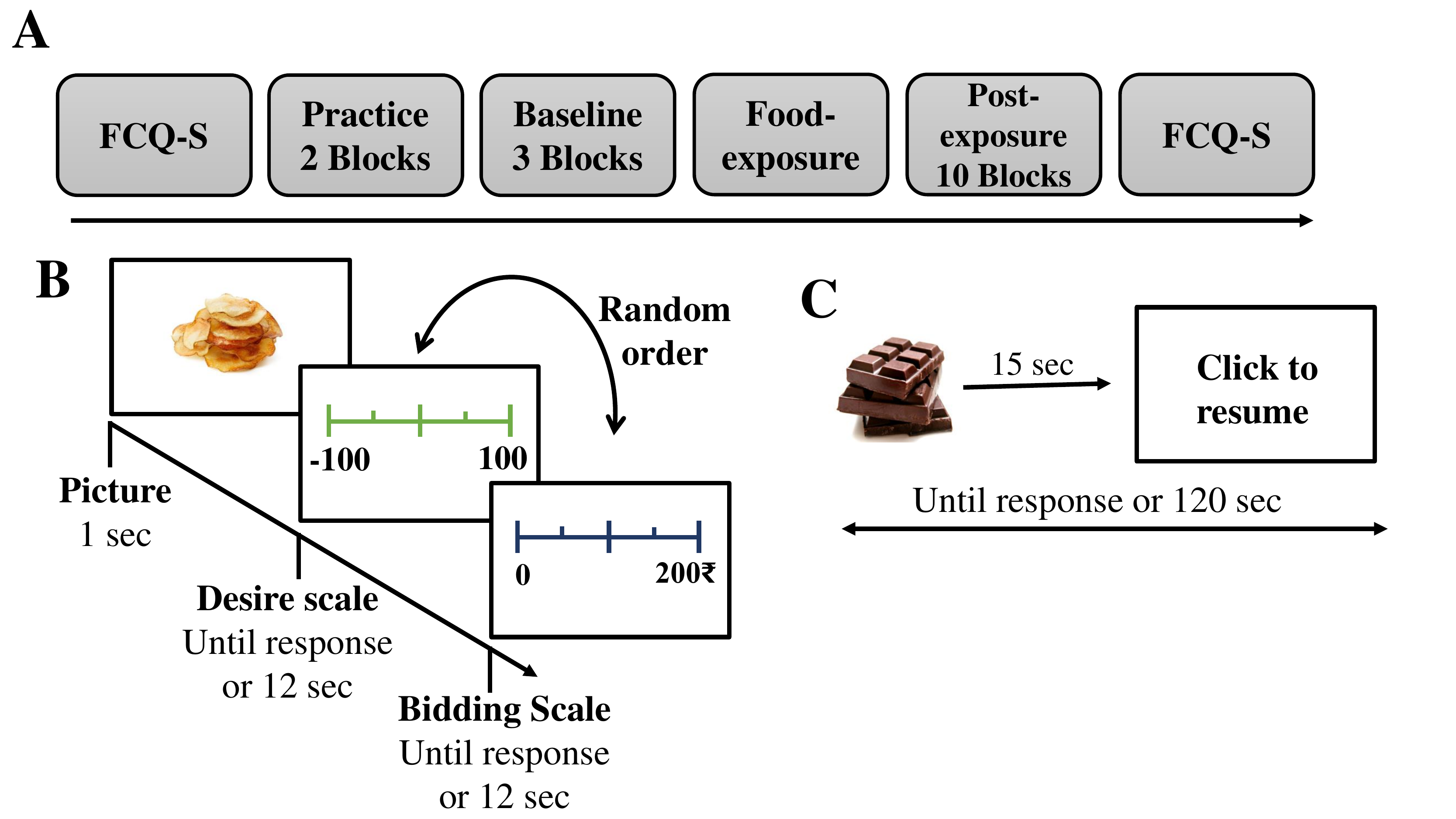}
			\caption{{\bf Experimental task.}
				A: Timeline of the Experimental procedure. B: Example of one trial consisting food item, desire scale and bidding scale. C: Maximum two minutes of multisensory exposure. Participants are asked to sense the food item but not to have it.}
			\label{fig1}
		\end{center}   
	\end{figure}

\subsubsection*{Procedure}
The behavioural data was collected for 57  (17  NC  +  19  PC  +  21  NEC) participants. Participants had to refrain from eating or drinking anything (except water) for at least two hours prior to the experiment to increase motivation for the sweet and savory food items offered. We noted time of their last meal at the beginning of the experiment. All participants entered their response in the Food Craving Questionnaire-State \cite{cepeda2000development,nijs2007modified} (FCQ-S; Fig~\ref{fig1}A) before starting and after completing the experiment. FCQ-S is a 15-item well founded  and definitive measure of state fluctuations in self-reported food craving. Responses for each of the questions were scored on a likert scale from 1 to 5 and a total state Food Craving score, which ranges from 15 to 75 was calculated. High scores indicate strong food craving experience and low scores indicate low food craving experience. The task included  a total of  15 blocks (2 practice  + 3 preexposure + 10 postexposure blocks; Fig~\ref{fig1}A). The first two practice blocks were not considered in the analysis of the experiment. Each block consisted of 10 trials (5 chocolate + 5 chips), leading to a total of 130 trials (excluding the practice trials). Each trial consisted of one high-resolution color image of a food item, a desire scale shown in green colour and a bidding scale in deep blue (Fig~\ref{fig1}B). Both the scales were mouse-controlled with a slider provided to help reach the required response. The desire scale ranges from the value -100 to 100, where -100 and 100 stand for the minimum and  maximum desire for the food item respectively. Participants were given a virtual rupee~200 endowment for each block, with which they could bid for the food items. Thus, the range of the bidding scale was dynamic and changed with each trial. It ranged from 0 to the amount left for rest of the trials (i.e., rupee~200 -amount spent till the trial) in a particular block. Desire and bidding scales were assigned in a random order in a trial so that participants could not predict the appearance of either of the scales. Participants were explicitly instructed to rate their subjective value in the form of a virtual monetary bid value and their concurrent desire for the displayed food item at that moment. Each participant experienced  multisensory food exposure of maximum of two minutes after completion of three preexposure blocks and before resuming the remaining 10 postexposure blocks (Fig~\ref{fig1}A) to induce food craving. Participants were given a real food item (chocolate or chips)  according to their class levels (i.e, NC, PC or NEC) and instructed to sense the food item by smelling and seeing but without having it. The whole experiment was monitored through a camera from outside the experiment room specially to confirm that  participants  refrained themselves from eating at the time of multisensory exposure. Participants received rupee~100 and a snack food item that he/she desires the most for their participation and a bonus amount, which was calculated using standard Becker-DeGroot-Marschak (BDM) mechanism \cite{becker1964measuring} with a minor modification (see Supporting Information for details).\\
\begin{figure}[!h]
	\begin{center}
		\includegraphics[width=16cm,height=9cm]{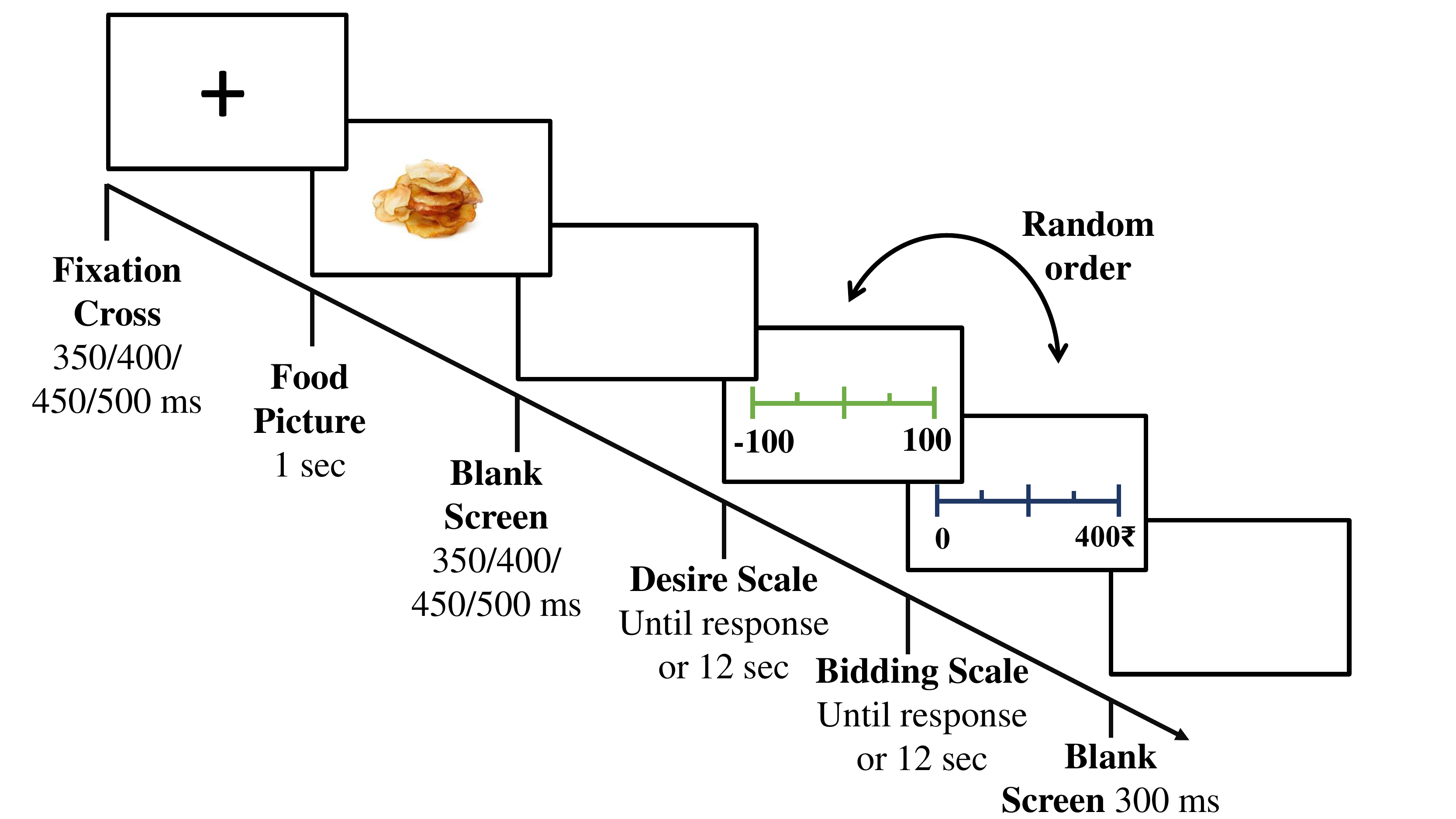}
		\label{fig2A}
		\caption{{\bf Example trial of EEG data collection.} A fixation screen with center cross was displayed  with a jittered time window (350-500 ms) before the appearance of food stimuli.  Two additional blank screens were  added. First one was after the stimuli was shown and the other one was at the end, i.e., after the 2nd response screen.
		}
	\end{center}   
\end{figure}

 In the second experiment EEG data was collected for 39 (13 NC + 13 PC + 13 NEC) participants. Data of four participants (1 NC + 1 PC +2 NEC) is excluded from the analysis due to faulty recording. The main experimental paradigm for EEG data collection remains the same. In the EEG experiments, 10 additional trials were used in each of the blocks, which resulted in 10 chocolate and 10 chips trials per block. Participants were given a virtual endowment of rupee~400  for each of the blocks for the bidding. Participant's compensation details are mentioned in the supporting information.

\subsection*{Neural Data Acquisition and Preprocessing}
EEG activity was recorded using 64-channel active shielded electrodes mounted in an EEG cap, which follows the international 10/20 system. We have used two linked Nexus-
32 bioamplifiers at a sampling rate of 512 Hz to record the EEG signals. Trials are epoched using the available trigger information and epoched trials are preprocessed using the EEGLAB toolbox \cite{delorme2004eeglab}. Trials were band-pass filtered (0.01 - 50 Hz.), baseline corrected and referenced using average referencing. Trials with ocular artifacts like blinks and eye movements were detected using bipolar electro-occulograms (EOG) and visual inspection and trials with amplitude exceeding $\pm$ 100 mV  were excluded from the analysis. $3.6\%$ of the trials were discarded as outliers and not used in the analysis.

\subsection*{Behavioral Data Analyses Techniques}
Inducement of cravings on participants was measured using the FCQ-S values before and after completion of the experiment.
In order to compare FCQ-S value between before start and after completion of the Neuroeconomic Decision-Making task, the total FCQ-S value is computed for each of the participants by adding the response value (between 1-5) for all 15 questions. Further, paired t-test is performed between before and after conditions for each of the control groups to check whether there is a significant increase in FCQ-S value at the end of the experiment.

Time dependency between participants' last meal taken and inducement of craving was checked using chi-square (${\chi}^2$) tests of independence based on a 2$\times$2 contingency table. The classes were formed in the following fashion. 
Time duration refrained from eating ($\geq$ 2 hours), before coming to the lab was reported by all the participants. Median of such time duration was calculated and based on the median value, two classes, shorter duration and longer duration were formed. Participants depending on their time from their last meal were sorted in the two classes (see details in the Supporting Information). 
We then calculated the difference ($d_i$) between FCQ-S values (after-before) for each participants. Participant $i$ is classified in YES Craving class if $d_i>\frac{s}{\sqrt{n}}t_{(n-1),\alpha}$ else in NO craving class, where $n=57$ (Number of participants), $s=$ standard deviation and $t_{(n-1),\alpha}$ is the $(1-\alpha)$th quantile of t-distribution with $(n-1)$ degrees of freedom. In our analysis we have chosen $\alpha$ = 0.05.

In order to compare the desire ratings as well as the bid-values (proportion of the bid amount out of the left amount), between the preexposure and postexposure blocks, two two-way ANOVA models with time (preexposure and postexposure) and control group (PC, NC and NEC) were used. 
The desire and bid-value for each trial were calculated after normalizing each participants' recorded desire and bid-values by the respective mean preexposure values for the exposed food item (i.e., either chocolate or chips trials, whichever was shown in the multi-sensory food exposure). Finally, the mean of the desire and bid-values for the exposed food item across the participants were calculated, which are considered as response variables of these models. 

 We  identified any change points present in the exposed time series (time series of desire and bid-values for exposed food item), i.e., we checked whether the means of these time series have any sudden change at any  point of time. Change point analysis \cite{killick2012optimal} tries to identify and estimate times when the probability distribution of a stochastic process or time series changes. We have used AMOC method \cite{killick2014changepoint} to detect change point for both bid value and desire value time series for the exposed food item.
 
 Using predictive models, we also computed the association between bid value, desire value and reaction time of bid value response. Regression models were computed keeping  bid value as response variable and taking recorded desire value and reaction time  as the regressor variables for the exposed food items for all control groups.
 While formulating models for NC, PC and NEC participants incorporating spline regression was needed as the respective scatter plots of bid value and desire rating indicate (see figure \ref{s3}). The main idea of a spline regression model is to divide the range of the $x$ (explanatory variable) into segments and fit appropriate spline functions in each of the segments. Splines are piece wise polynomials of order $k$ and the joint points of the segments are generally termed as knots.  In our study, knots of SRM were determined following results of scatter plot (\ref{s3}). The general spline regression model is defined in the Supporting Information. In order to reject the outliers from the data, we have used Cook's distance method \cite{cook1977detection} for each of the models.
 Details of our model can be found in the Results section.
 
 \subsection*{Neural Data Analyses Techniques}
 In order to find out the differential ERP amplitude between preexposure and postexposure conditions for the exposed food item, we investigated  P200, N200, P300 and Late Positive Potential (LPP) for all control participants.  P200 was measured at the frontal cluster (F1, F2, F3, F4, Fz, AF3, AF4) around 200 ms after stimulus onset.
 N200 was measured at the parietal electrode cluster (P1, P2, P3, P4, Pz, PO3, PO4) in an early time window around 200 ms after stimulus onset. P300 and LPP were measured at the centro-parietal electrode cluster (CP1, CP2, CP3, CP4, CPz, P1, P2) around 300 ms and 300 to 600 ms time window respectively. Peak amplitudes were identified for all the ERP components except LPP. In LPP, the mean amplitude  in the 300 to 600 ms time window is measured. To compare the ERP components between preexposure and postexposure conditions,  one-way repeated measure ANOVAs were computed for each of the control groups.
 
 Partial correlations were calculated for each group at both preexposure and postexposure conditions to find out the relation between dependent measures, i.e., between subjective desire ratings and ERP components (P200, N200, P300 and LPP). Alike other studies \cite{wolz2017subjective,kelley2012effect,cohen2013statistical}, $r > 0.24$ (corresponds to $d>0.5$) and  $r > 0.3$ (corresponds to $d>0.8$) were considered as moderate and large correlation, respectively.
 
 Univariate EEG analysis is popular and widely used to establish the relationship between behavioral performance and neural activity in several cognitive tasks. However, the univariate EEG analysis techniques fail to  completely employ the spatio-temporal structure of the multivariate neural data. Applications of multivariate pattern analysis (MVPA) techniques help to incorporate the spatial and temporal information present in the EEG data by integrating the neural information into a single decision variable. Successful application of MVPA has been depicted in numerous studies using EEG and fMRI (\cite{haynes2005predicting,philiastides2006neural,das2010predicting}). 
 Since EEG data is high dimensional and often suffers from the small sample size problem we have used Classwise Principal Component Analysis (CPCA) \cite{das2009efficient} to classify preexposure and postexposure classes for all control participants.
 CPCA, a supervised MVPA technique has been successfully used in previous studies \cite{das2009mental,das2010predicting,do2011brain,do2013brain,saha2020our,roy2021wisdom} to reduce the dimensionality of the EEG signals and extract informative features. The technique is based on the application of principal component analysis (PCA) in each of the classes and aims to identify and discard the non-informative subspace present in the data. The classification is then carried out in the residual space in which small sample size problems and the curse of dimensionality no longer hold. Classification for single-trial EEG data  was computed using Linear Bayesian Classifier. Pattern analysis was performed using leave one trial out cross validation method for each of the individual participants and mean classification accuracies for all controls are reported in the results. Further, paired t-test is performed to check whether the classification accuracy is significantly above chance.

 \section*{Results}
 \subsection*{FCQ-S value}
 
\begin{figure}[!h]
 	\begin{center}
 		\includegraphics[width=15cm,height=8cm]{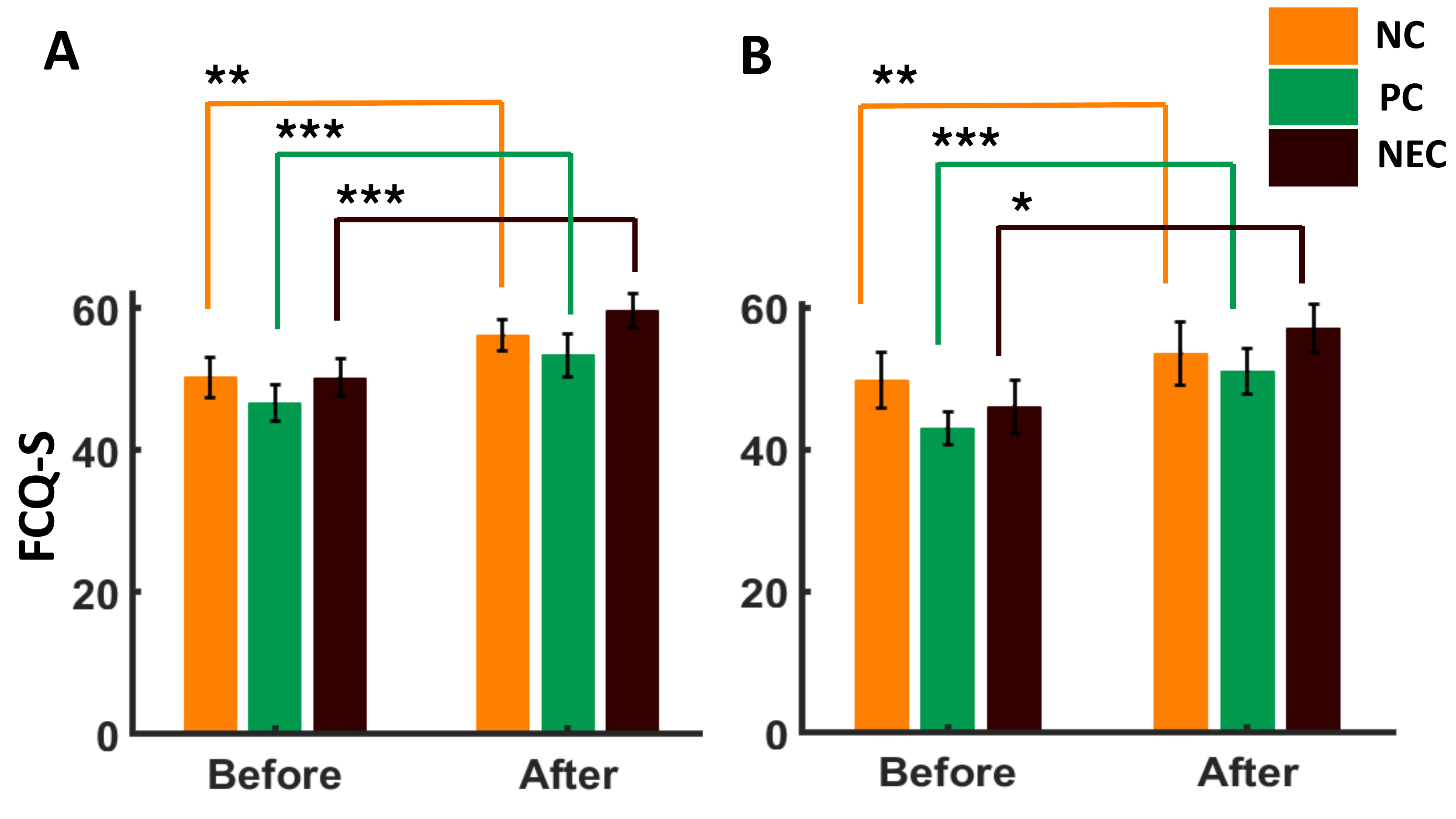}
 		\caption{\textbf{FCQ-S values (means $\pm${~SEM}) at before start of and after completion of the experiment.} A. FCQ-S values for the behavioural experiment. B. FCQ-S values for EEG experiment. The figure depicts an increment in FCQ-S value after completion of the experiment for all control groups in both the experiments.   ${}^{*}p < 0.05,{}^{**}p < 0.01, {}^{***}p < 0.001$. }
 		
 		\label{fig3}
 	\end{center}   
 \end{figure}

 We have performed paired-sample $t$-test for all control groups, to compare measured FCQ-S values at before start of and after completion of the experiment. In all the control groups, FCQ-S value increased after completion of the experiment
 (Figure~\ref{fig3}A), indicating an inducement in food craving like state throughout the experiment. The increment in mean FCQ-S value was found to be significant for all groups (NC: $t_{16}=2.71$, $p-$value $= 0.0077$, PC: $t_{18}=4.34$,  $p-$value $= 1.9690\times 10^{-4}$ and  NEC: $t_{20}=4.56$, $p-$value $= 9.4735\times 10^{-5}$). Similarly, a significant increment in FCQ-S value after completion of the experiment is observed for all control groups performing the EEG experiment (see Figure~\ref{fig3}B), NC: $t_{11}=3.18$, $p-$value $= 0.0044$, PC: $t_{11}=4.03$,  $p-$value $= 9.91\times 10^{-4}$ and  NEC: $t_{10}=2.52$, $p-$value $= 0.0151$).
 
 \begin{table}[!tbp]
 	\begin{center}
 		\setlength{\arrayrulewidth}{.3mm}
 		\begin{tabular}{lccc}
 			\hline
 			\multicolumn{1}{l}{}&\multicolumn{1}{c}{Shorter duration}&\multicolumn{1}{c}{Longer duration}&\multicolumn{1}{c}{Test Statistic}\tabularnewline
 			&\multicolumn{1}{c}{{\scriptsize $N=29$}}&\multicolumn{1}{c}{{\scriptsize $N=28$}}&\tabularnewline
 			\hline
 			~~~~YES craving&66\%~{\scriptsize~(19)}&75\%~{\scriptsize~(21)}&$ \chi^{2}(1)=0.6121 ,~ P=0.43  $\tabularnewline
 			~~~~NO craving&34\%~{\scriptsize~(10)}&25\%~{\scriptsize~(7)}&\tabularnewline
 			\hline
 			
 		\end{tabular}
 		\caption{Time duration difference in response to question of whether food cravings had ever been
 			experienced}
 		\label{time_duration}
 	\end{center}
 	
 \end{table}

 \subsection*{Time Duration Refrained from Eating}
The chi-square test to check the effect of time duration since last meal on craving reveals (table \ref{time_duration}) that there is no reason to believe that craving is dependent on the time since last meal ($\chi^{2}(1)=0.6121 ,P=0.43$). Similarly, 
the chi-square test for the EEG experiment shows the same result ($\chi^{2}(1)=0.2292 ,~ P=0.6321$).

\subsection*{Behavioural Data Analysis}
Two-way ANOVA on bid-value resulted in a significant main effect of time factor ($F_{1,189}=23.71, P = 0.000$) and control group factor ($F_{2,189}=6.74, P = 0.0015$) and a significant interaction (time $\times$ control) ($F_{2,189}=6.75, P = 0.0015$). Multiple comparison tests followed by ANOVA was performed and a significant  difference in bid- values between time levels only for the NC participant was observed (Table: \ref{table1}). Estimated mean difference being negative for the NC group (Table: \ref{table1}) showed higher willingness to pay in the postexposure blocks than the preexposure blocks. 
Similarly, two-way ANOVA for the desire ratings followed by multiple comparisons between group means was performed.
A significant main effect, on time factor ($F_{1,189}=59.87, P = 5.92 \times 10^{-13}$), control group factor ($F_{2,189}=37.85, P = 1.49 \times 10^{-14}$) and a significant interaction (time $\times$ control) ($F_{2,189}=14.03, P = 2.09 \times 10^{-06}$)  was noticed. Multiple comparison tests confirmed the desire value in the postexposure blocks increased significantly than the preexposure blocks for both the NC and NEC groups (Table: \ref{table1}). Hence, our results show that for both NC and NEC groups, desire for exposed food items increased significantly compared to preexposure however, only participants from NC group were willing to pay more for exposed food items.

\begin{table}
	\centering
	\renewcommand{\arraystretch}{1.2}
	\setlength{\arrayrulewidth}{.3mm}
	\rowcolors{4}{}{}
	\begin{tabular}{|p{3.2cm}|c|c|c|c|c|}
		
		\hline
		\multirow{2}{5cm}{\textbf{Control Group}} & \multicolumn{2}{c|}{\textbf{$\mathbf{\hat{\mu}_{Pre}-\hat{\mu}_{Post}}$}} & \multicolumn{2}{c|}{\textbf{$p$- value}}\\
		\cline{2-5}
		& \cellcolor{gray!60!yellow!70!}\textbf{Bid Value} &\cellcolor{gray!60!yellow!70!} \textbf{Desire Value} & \cellcolor{gray!60!yellow!70!}\textbf{Bid Value} &\cellcolor{gray!60!yellow!70!} \textbf{Desire Value}\\
		\hline
		\bf Positive (PC) & $-0.2388$ & $-0.0259$ & $0.9253$ & $1.000$ \\ \hline
		\bf Negative (NC) & $-1.4134$ & $-1.2043$ &\cellcolor{green!25!blue!35} $0.0000$*** &\cellcolor{green!25!blue!35} $0.0000$*** \\ \hline
		\bf  Neutral (NEC) & $-0.4083$& $-1.2096$ & $0.5511$&\cellcolor{green!25!blue!35} $0.0000$*  \\ \hline
		
	\end{tabular}
	\caption{Multiple comparison of group means on bid and desire value. $\mathbf{\hat{\mu}_{Pre}} = $Estimated mean at preexposure;  $\mathbf{\hat{\mu}_{Post}} =$ Estimated mean at postexposure. Control groups with blue highlighted p-values ($< 0.05$) are significant. ${}^{*}p < 0.05,{}^{**}p < 0.01, {}^{***}p < 0.001$}
	\label{table1}
\end{table}

\subsection*{Change Point Detection}
\begin{figure}[!h]
	\begin{center}
		\includegraphics[width=17cm,height=10cm]{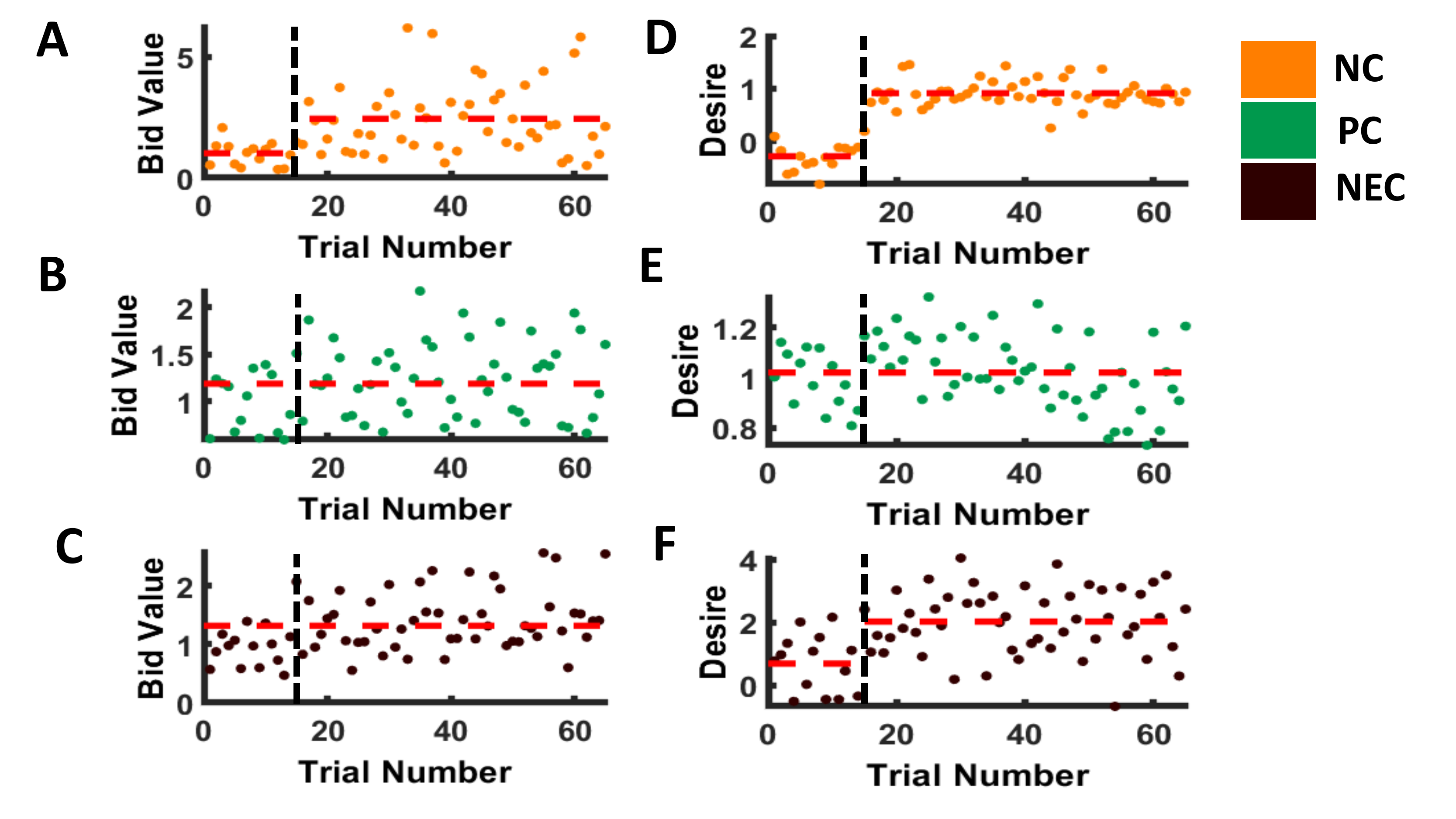}
		\caption{{\bf Change point detection for exposed series.}
			Scatter plot of all controls for both bid value and desire value exposed foods. Vertical black dashed lines indicate trial 15, after which multisensory food exposure has occurred. Red dashed lines indicate estimated mean value and a jump in that line indicates the occurrence of a change point. This clearly depicts that both bid value and desire value exposed series showed a single change point at  sixteenth and  fifteenth trail respectively for NC, no change points for PC and only one change point at fourteenth trial for NEC desire value exposed series.} 
		\label{fig2}
	\end{center}   
\end{figure}

Bid-value series of the NC group showed one single change point at the sixteenth trial (the first trial of postexposure)  with the estimated bid value  mean $1.0247$ and $2.4352$ respectively before and after the change point (Fig~\ref{fig2}A). But no change point was detected, with the estimated bid value mean $1.1851$ and $1.3153$  for both PC and NEC bid value series, respectively (Fig~\ref{fig2}B and Fig~\ref{fig2}C). Similarly, the desire value series of the NC group showed one single change point at the fifteenth trial with the estimated mean desire value $-0.2941$ and $0.9102$ respectively before and after the change point (Fig~\ref{fig2}D) and no change point was detected for PC desire series with $1.0199$ as the estimated mean (Fig~\ref{fig2}E). But unlike the bid value series, desire value series for the NEC group showed one single change point at the fourteenth trial, with the estimated mean $0.6946$ and $2.0269$, respectively, before and after the occurrence of the change point (Fig~\ref{fig2}F). Thus, results of change point detection are concurrent with our 2 way ANOVA results described in the previous section. {Further, change point detection analysis on desire values of non exposed food items shows a single change point at twelfth trial only for NC with estimated desire values $1.1669$ and $0.0667$, respectively, before and after the occurrence of the change point (see Figure \ref{s4}), thus showing reduction in desire for previously liked food. Hence our paradigm clearly indicates a reversal effect of food preference after multisensory food exposure.}     

\subsection*{Computational Model}

\begin{figure}[!h]
	\begin{center}
		\includegraphics[width=16cm,height=10cm]{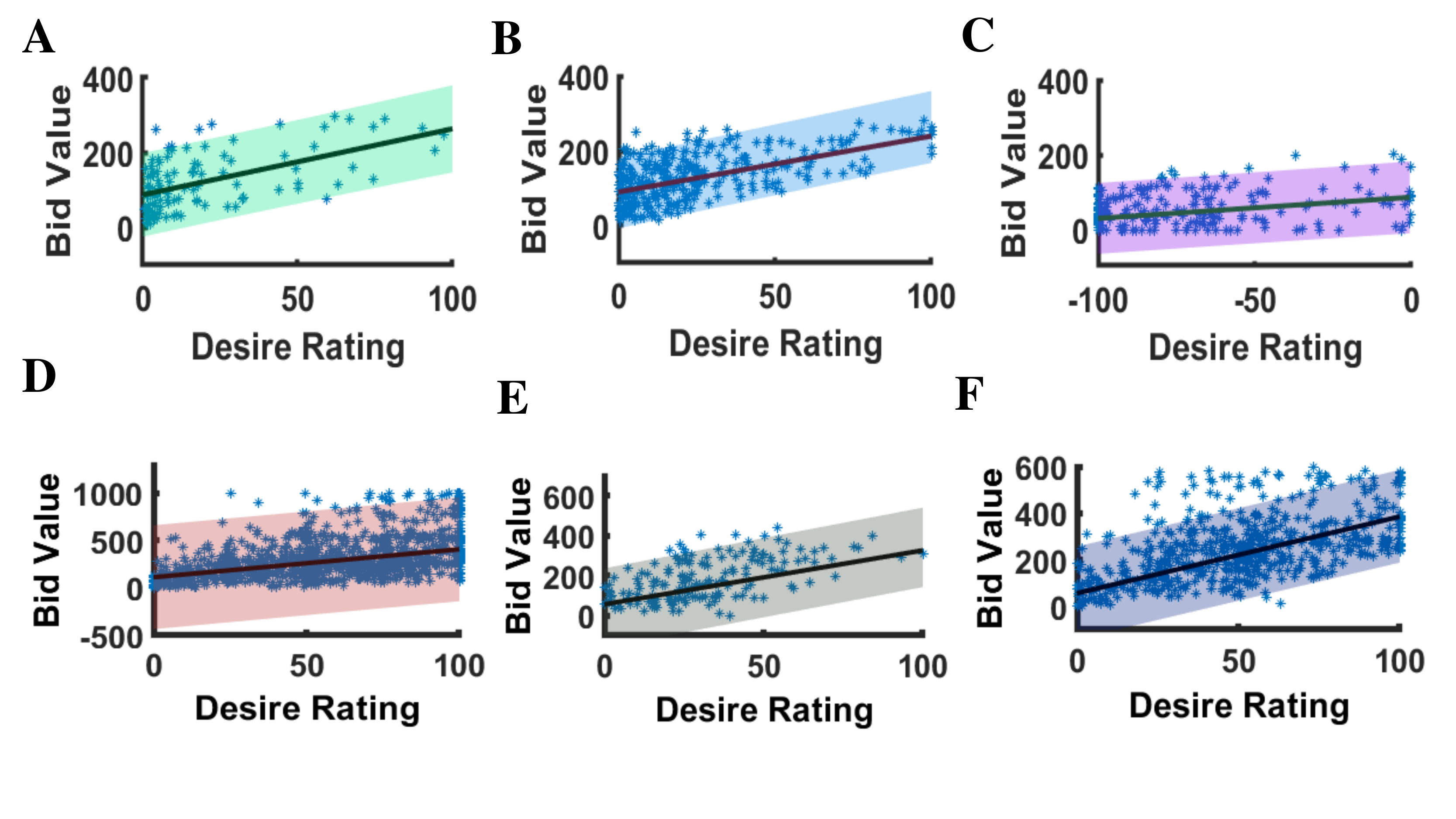}
		\caption{{\bf Model prediction of Bid Value from Desire Rating.}
			A: NC preexposure model when the desire rating is with in the interval [0,100]. B: NC postexposure model when the desire rating is with in the interval [0,100]. C: NC postexposure model when the desire rating is with in the interval [-100,0). D: PC model when the desire rating is with in the interval [0,100]. E: NEC model at preexposure. F: NEC model at postexposure. Blue dots are the data points and the straight line along with the colored shaded region indicate SRM prediction with 95\% confidence intervals. Almost all the data points are within the 95\% confidence interval, which shows the SRM fits the data well for all control participants.}
		\label{fig4}
	\end{center}   
\end{figure}

\textbf{Negative Control (NC) :} The two-way ANOVA model and change point detection analysis confirmed a significant change in mean value of preexposure compared to the postexposure blocks for both  bid value and desire value exposed series.  So, we have considered two distinct spline regression models (SRMs) for preexposure and postexposure blocks. 
Our preexposure \eqref{eq:1} and postexposure \eqref{eq:2} models can be written as follows: 
\begin{equation}
	BV_{\mbox{\tiny{Pre}}}  = ({\alpha_{\mbox{\tiny{0}}}} + {\alpha_{\mbox{\tiny{1}}}} DV_{\mbox{\tiny{Pre}}})I (DV_{\mbox{\tiny{Pre}}} < 0) + (\alpha_{\mbox{\tiny{2}}} + {\alpha_{\mbox{\tiny{3}}}} DV_{\mbox{\tiny{Pre}}})I (DV_{\mbox{\tiny{Pre}}} \geq  0) + \alpha_{\mbox{\tiny{4}}} RT_{\mbox{\tiny{Pre}}} + \epsilon
	\label{eq:1}
\end{equation}
\begin{equation}
	BV_{\mbox{\tiny{Post}}}  = ({\beta_{\mbox{\tiny{0}}}} + {\beta_{\mbox{\tiny{1}}}} DV_{\mbox{\tiny{Post}}})I (DV_{\mbox{\tiny{Post}}} < 0) + (\beta_{\mbox{\tiny{2}}} + {\beta_{\mbox{\tiny{3}}}} DV_{\mbox{\tiny{Post}}})I (DV_{\mbox{\tiny{Post}}} \geq 0) + \beta_{\mbox{\tiny{4}}} RT_{\mbox{\tiny{Post}}} + \epsilon,
	\label{eq:2}
\end{equation}
where $BV_i$, $DV_i$ $\&$ $RT_i$ denote the bid value, desire value and reaction time of bid value, respectively, at preexposure or at postexposure depending on $i = Pre$ or $Post$, and $\alpha$'s and $\beta$'s are the coefficients of the model \eqref{eq:1} and \eqref{eq:2} respectively. Preexposure model turned out to be significant ($F_{5,232} = 142.1$, $ P < 2.2 \times 10^{-14}$, Adjusted R - squared $= 0.75$) and all but ${\alpha_{\mbox{\tiny{1}}}}$ and  ${\alpha_{\mbox{\tiny{4}}}}$ were significant (see  Table: \ref{table3}). {To check the predictive performance of our proposed model, we estimated the predictive band and depicted in Figure~\ref{fig4}A. As clearly seen, majority of the true values lie within the 95\% predictive bands.}

\begin{center}
	\begin{table}[!ht]
		\begin{center}
			\setlength{\arrayrulewidth}{.3mm}
			\setlength{\tabcolsep}{13pt}
			\renewcommand{\arraystretch}{1}
			{\rowcolors{3}{gray!60!yellow!40}{gray!60!yellow!40}
				
				\begin{tabular}{ |p{2.5cm}|p{3cm}|p{2.5cm}|p{3cm}|  }
					\hline
					\multicolumn{4}{|c|}{\bf Spline regression  model (NC preexposure). } \\
					\hline
					\bf Coefficients &    \bf Estimate  & \bf t-value & \bf p-value \\
					\hline
					${\alpha_{\mbox{\tiny{0}}}}$ &  $ 40.54$ &$2.638$ &\cellcolor{green!25!blue!35} $0.0089$ ** \\
					${\alpha_{\mbox{\tiny{1}}}}$ &  $ -0.20$ &$-1.193$ &  $0.2342$ \\
					${\alpha_{\mbox{\tiny{2}}}}$ &  $87.619$ &$8.066$ &\cellcolor{green!25!blue!35} $3.88 \times 10^{-14}$ *** \\
					${\alpha_{\mbox{\tiny{3}}}}$ &  $1.7617$ &$7.646$ &\cellcolor{green!25!blue!35} $5.47 \times 10^{-13}$ *** \\
					${\alpha_{\mbox{\tiny{4}}}}$ &  $0.0002$ &$0.098$ & $0.9218$ \\

					\hline
				\end{tabular}
			}
			\caption{
				{\bf Spline regression model results.} Coefficients with highlighted p-values are significant in the model.
				${}^{*}p < 0.05,{}^{**}p < 0.01, {}^{***}p < 0.001$}
			\label{table3}
		\end{center}
	\end{table}
\end{center}
Also the postexposure model appeared to be significant ($F_{5,560} = 560.1$, $ P < 2.2 \times 10^{-16}$, Adjusted R - squared =0.83) and all the coefficients except $\beta_{\mbox{\tiny{4}}}$ had significant effect in the model (see Table: \ref{table4}). {The predictive performance of our proposed model is checked using the predictive bands and are given in Figure~\ref{fig4}B and ~\ref{fig4}C. Most of the true values lie within the 95\% predictive bands.} 
\begin{center}
	\begin{table}[!ht]
		\begin{center}
			\setlength{\arrayrulewidth}{.3mm}
			\setlength{\tabcolsep}{13pt}
			\renewcommand{\arraystretch}{1}
			{\rowcolors{3}{gray!60!yellow!40}{gray!60!yellow!40}
				
				\begin{tabular}{ |p{2.5cm}|p{3cm}|p{2.5cm}|p{3cm}|  }
					\hline
					\multicolumn{4}{|c|}{\bf Spline regression  model (NC postexposure). } \\
					\hline
					\bf Coefficients &    \bf Estimate  & \bf t-value & \bf p-value \\
					\hline
					${\beta_{\mbox{\tiny{0}}}}$ &  $ 86.4553$ & $8.693$ &\cellcolor{green!25!blue!35} $< 2 \times 10^{-16}$ *** \\
					${\beta_{\mbox{\tiny{1}}}}$ &  $ 0.5655$ &$5.43$ & \cellcolor{green!25!blue!35} $8.42 \times 10^{-8}$ *** \\
					${\beta_{\mbox{\tiny{2}}}}$ &  $89.8364$ &$15.906$ &\cellcolor{green!25!blue!35} $< 2 \times 10^{-16}$ *** \\
					${\beta_{\mbox{\tiny{3}}}}$ &  $1.4947$ &$13.273$ &\cellcolor{green!25!blue!35} $< 2 \times 10^{-16}$ *** \\
					${\beta_{\mbox{\tiny{4}}}}$ &  $0.001734$ &$1.004$ & $0.316$ \\

					\hline
				\end{tabular}
			}
			\caption{
				{\bf Spline regression model results.} Coefficients with highlighted p-values are significant in the model.${}^{*}p < 0.05,{}^{**}p < 0.01, {}^{***}p < 0.001$ }
			\label{table4}
		\end{center}
	\end{table}
\end{center}
No effect of reaction time on bid value was observed for both preexposure and postexposure models. 
In the NC preexposure model, the non significant  $\alpha_{\mbox{\tiny{1}}}$ shows no dependency between bid value and desire value for the non desired food items. But $\beta_{\mbox{\tiny{1}}}$ was significant in the NC postexposure model, which implies a effect of desire value on bid value for the non desired food items postexposure.

\textbf{Positive Control (PC) :} The two-way ANOVA model and the change point detection analysis confirmed no significant change in the mean value at preexposure compared to the postexposure blocks for both  bid value and desire value of exposed series, which drove us to consider only one spline regression  model for the entire exposed series. Alike the NC models, knot was taken at desire zero. Our proposed PC model \eqref{eq:3} can be written as follows:
\begin{equation}
	BV = ({\gamma_{\mbox{\tiny{0}}}} + {\gamma_{\mbox{\tiny{1}}}} DV)I (DV < 0) + (\gamma_{\mbox{\tiny{2}}} + {\gamma_{\mbox{\tiny{3}}}} DV)I (DV \geq  0) + \gamma_{\mbox{\tiny{4}}} RT + \epsilon
	\label{eq:3}
\end{equation}
where $BV$, $DV$ $\&$ $RT$ denote the bid value, desire value and reaction time of bid value, respectively,  for the PC group and $\gamma$'s are the coefficients of the model. This model turned out to be significant ($F_{5,1082} = 1065$, $ P < 2.2 \times 10^{-16}$, Adjusted R - squared = 0.83) and all but ${\gamma_{\mbox{\tiny{1}}}}$ and  ${\gamma_{\mbox{\tiny{4}}}}$ were significant (see  Table: \ref{table5}). {Here also, like in negative control studies, we check the predictive performance of our model by estimating the 95\% predictive band. The band along with the true values are displayed in Figure~\ref{fig4}D, which clearly showed the band captures most of the true values. This indicates that our proposed model is not suffering from overfitting.}
\begin{center}
	\begin{table}[!ht]
		\begin{center}
			\setlength{\arrayrulewidth}{.3mm}
			\setlength{\tabcolsep}{13pt}
			\renewcommand{\arraystretch}{1}
			{\rowcolors{3}{gray!60!yellow!40}{gray!60!yellow!40}
				
				\begin{tabular}{ |p{2.5cm}|p{3cm}|p{2.5cm}|p{3cm}|  }
					\hline
					\multicolumn{4}{|c|}{\bf Spline regression  model (PC). } \\
					\hline
					\bf Coefficients &    \bf Estimate  & \bf t-value & \bf p-value \\
					\hline
					${\gamma_{\mbox{\tiny{0}}}}$ &  $ 55.63$ & $2.062$ &\cellcolor{green!25!blue!35} $0.0394$ * \\
					${\gamma_{\mbox{\tiny{1}}}}$ &  $ -0.0699$ &$-0.041$ &  $0.9672$ \\
					${\gamma_{\mbox{\tiny{2}}}}$ &  $113.9$ &$10.823$ &\cellcolor{green!25!blue!35} $< 2 \times 10^{-16}$ *** \\
					${\gamma_{\mbox{\tiny{3}}}}$ &  $2.95$ &$21.751$ &\cellcolor{green!25!blue!35} $< 2 \times 10^{-16}$ *** \\
					${\gamma_{\mbox{\tiny{4}}}}$ &  $-0.0003$ &$-0.098$ & $0.9218$ \\

					\hline
				\end{tabular}
			}
			\caption{
				{\bf Spline regression model results.} Coefficients with highlighted p-values are significant in the model.${}^{*}p < 0.05,{}^{**}p < 0.01, {}^{***}p < 0.001$ }
			\label{table5}
		\end{center}
	\end{table}
\end{center}
In the PC  model, the non significant  $\gamma_{\mbox{\tiny{1}}}$ shows no effect of desire value on bid value  in the negative desire range and reaction time shows no effect in the model. However, the model depicts  a significant effect of desire value on bid value in the non negative desire range.

\textbf{Neutral Control (NEC) :} The two-way ANOVA model and the change point detection analysis confirmed no significant change in mean bid-value at preexposure compared to the postexposure blocks  but a significant change in mean desire value at preexposure compared to the postexposure blocks for the exposed series. Thus, we have considered a spline regression  model, with a knot at fifteenth trial for the entire exposed series. The proposed NEC model \eqref{eq:4} can be written as follows: 
\begin{equation}
	BV = {\delta_{\mbox{\tiny{0}}}} + ({\delta_{\mbox{\tiny{1}}}}DV)I(t\leq 15)) + ({\delta_{\mbox{\tiny{2}}}}DV)I(t > 15)) + \delta_{\mbox{\tiny{3}}}RT + \epsilon,
	\label{eq:4}
\end{equation}
where $BV$, $DV$,  $RT$ $\&$ $t$ denote the bid value, desire value,  reaction time of bid value and trial number respectively,  for the NEC group and $\delta$'s are the coefficients of the model. This model turned out to be significant ($F_{3,1185} = 581.9$, $ P < 2.2 \times 10^{-16}$, Adjusted R - squared = 0.60) and all the coefficients were significant (see  Table: \ref{table6}) in the model. {To nullify the possibility of overfitting, like in negative control and positive control studies, the predictive bands for preexpossure and postexpossure models are estimated. The 95\% bands and the true values are depicted in Figure~\ref{fig4}E and \ref{fig4}F. As expected, the major portion of the true values fell within the bands. }
\begin{center}
	\begin{table}[!ht]
		\begin{center}
			\setlength{\arrayrulewidth}{.3mm}
			\setlength{\tabcolsep}{13pt}
			\renewcommand{\arraystretch}{1}
			{\rowcolors{3}{gray!60!yellow!40}{gray!60!yellow!40}
				
				\begin{tabular}{ |p{2.5cm}|p{3cm}|p{2.5cm}|p{3cm}|  }
					\hline
					\multicolumn{4}{|c|}{\bf Spline regression  model (NEC). } \\
					\hline
					\bf Coefficients &    \bf Estimate  & \bf t-value & \bf p-value \\
					\hline
					${\delta_{\mbox{\tiny{0}}}}$ &  $ 26.6446$ & $5.506$ &\cellcolor{green!25!blue!35} $4.5 \times 10^{-8}$ *** \\
					${\delta_{\mbox{\tiny{1}}}}$ &  $ 2.7018$ &$13.084$ & \cellcolor{green!25!blue!35}  $< 2 \times 10^{-16}$ *** \\
					${\delta_{\mbox{\tiny{2}}}}$ &  $3.2891$ &$35.381$ &\cellcolor{green!25!blue!35} $< 2 \times 10^{-16}$ *** \\
					${\delta_{\mbox{\tiny{3}}}}$ &  $0.0198$ &$11.009$ &\cellcolor{green!25!blue!35} $< 2 \times 10^{-16}$ *** \\

					\hline
				\end{tabular}
			}
			\caption{
				{\bf Spline regression model results.} Coefficients with highlighted p-values are significant in the model. ${}^{*}p < 0.05,{}^{**}p < 0.01, {}^{***}p < 0.001$ }
			\label{table6}
		\end{center}
	\end{table}
\end{center}
The NEC model depicts a significant effect of desire value on bid value for the exposed food items throughout the experiment
and the association between bid value and desire value increased slightly after multisensory food exposure (since $\hat{\delta_{\mbox{\tiny{2}}}}-\hat{\delta_{\mbox{\tiny{1}}}}=0.5873$).  

\begin{figure}[!h]
	\begin{center}
		\includegraphics[width=13cm,height=19cm]{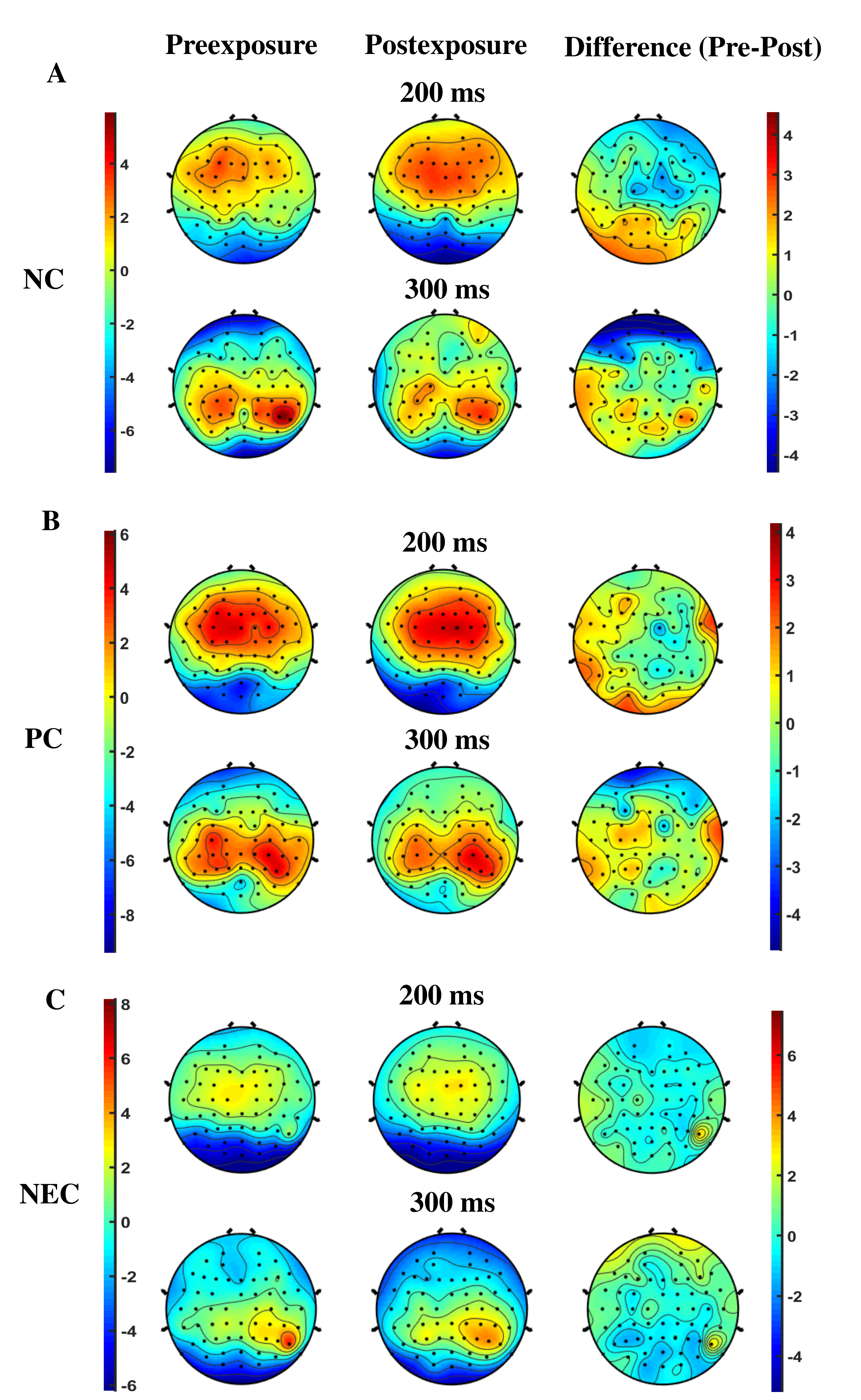}
		\caption{{\bf Topoplots  for preexposure, postexposure and the their differences (preexposure-postexposure).}
			Topoplots are depicted for A: NC, B: PC and C: NEC. at 200ms and 300ms. The colorbars at left side are for the reference of both preexposure and postexposure, whereas colorbars at right side are for the reference  for the differences. }
		\label{fig_topo}
	\end{center}   
\end{figure}

\subsubsection*{Model Summary}
\label{model summery}
Here we summarize the main findings of the behavioural model. 
\begin{enumerate}
	\item Our computational model clearly shows that for exposed food items, willing to pay (measured by bid value) can be predicted successfully from desire values for all category of participants. 
	\item For NC, our model shows insignificant preference for disliked food preexposure which is overturned and became significant following multisensory exposure. 
\end{enumerate}

\subsection*{Univariate EEG Analysis}
{To determine the timing and location of the maximum amplitudes, we have observed the topographic plots of grand averages at different time points for both preexposure and postexposure conditions using EEGLAB Toolbox \cite{delorme2004eeglab}.
	Further, to elucidate whether the preexposure and postexposure conditions induce different neural processing mechanisms, the grand average difference waveform was observed. A clear difference in preexposure and postexposure ERPs (N200 at 200ms and P300 at 300 ms) was visible only for NC (see Figure \ref{fig_topo}).}

\subsubsection*{P200}
\begin{figure}[!h]
	\begin{center}
		\includegraphics[width=18cm,height=13cm]{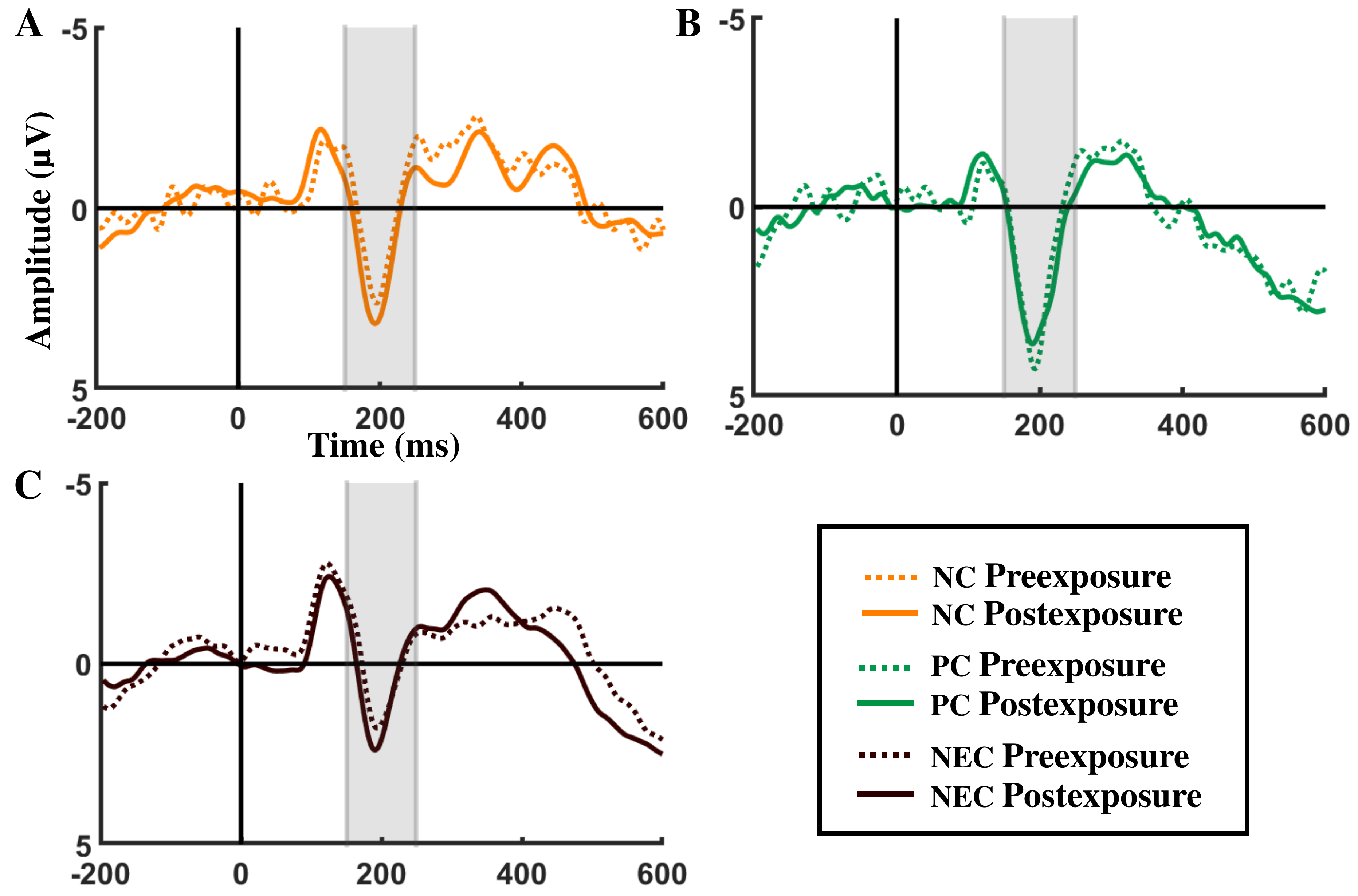}
		\caption{{\bf Event-related potentials in response to the food pictures (exposed) of a frontal cluster (F1, F2, F3, F4, Fz, AF3, AF4) for both preexposure and postexposure conditions.} The graph depicts the electrophysiological activity from 200 ms before to 600 ms after stimulus onset. A. ERPs for NC participants. B. ERPs for PC participants C.  ERPs for NEC participants. The grey shaded region indicates the time window, when the positive going peak is identified. There seems to be no difference in P200 amplitude between preexposure and postexposure for all the controls.  
		}
		\label{frontal_P200}
	\end{center}   
\end{figure}

We have observed a positive going peak around 200 ms after stimulus onset for all control groups (see Figure \ref{frontal_P200}). However, there was no significant main effect of time factor (i.e, preexposure vs postexposure) for any of the control groups (NC: $F(1,11)=0.5831, p= 0.4612$, PC: $F(1,11)=0.2029, p= 0.6612$ and NEC: $F(1,10)=0.0029, p= 0.9582$). Further, any control groups showed no localization effect (left vs right) of P200 amplitude at preexposure and postexposure (see Table \ref{localization_P200} for details).  

\subsubsection*{Posterior N200}
\begin{figure}[!h]
	\begin{center}
		\includegraphics[width=18cm,height=13cm]{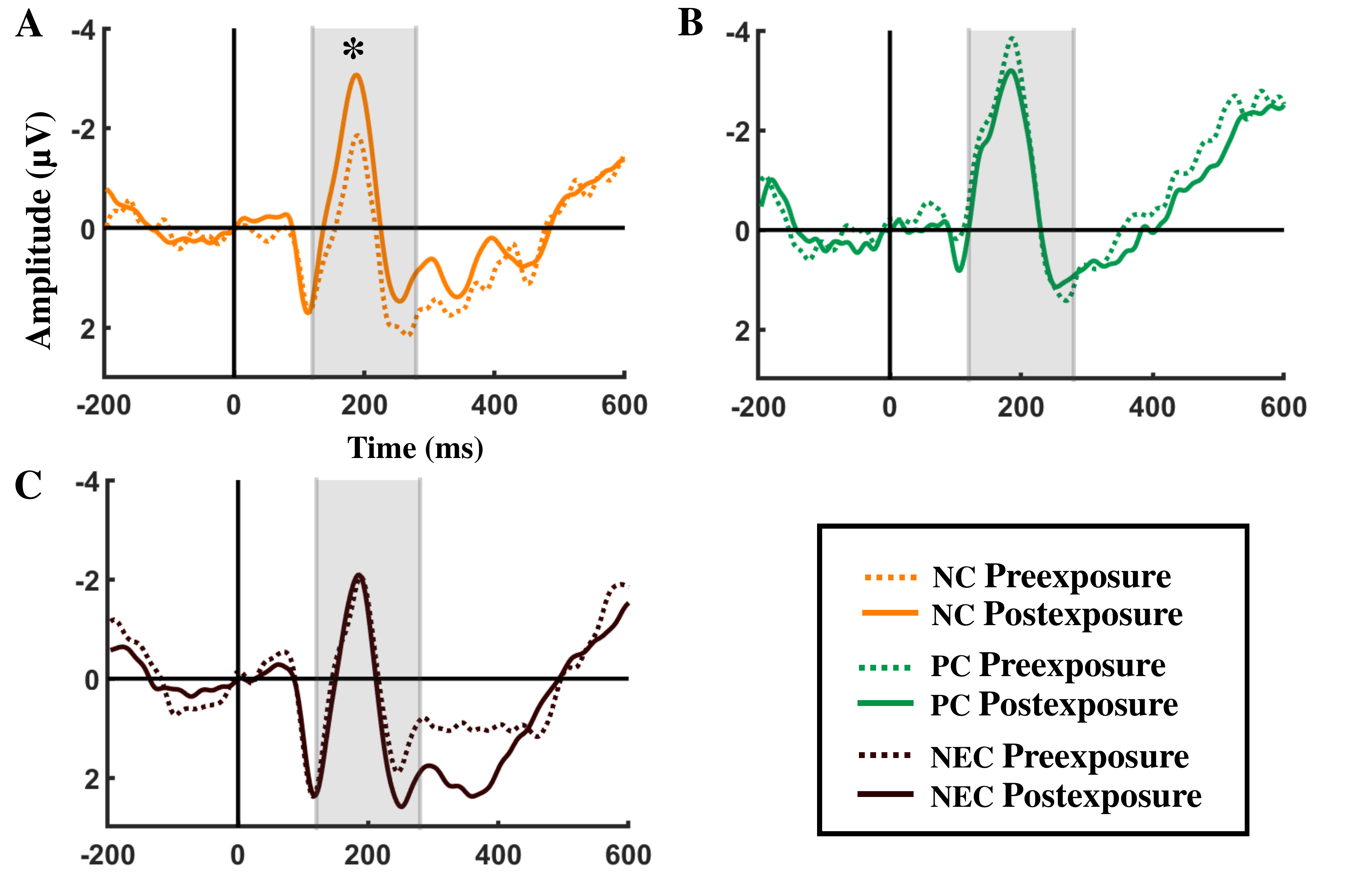}
		\caption{{\bf Event-related potentials in response to the food pictures (exposed) at  parietal cluster (P1, P2, P3, P4, Pz, PO3,  PO4) for both preexposure and postexposure conditions.} The graph depicts the electrophysiological activity from 200 ms before to 600 ms after stimulus onset. A. ERPs for NC participants. B. ERPs for PC participants C.  ERPs for NEC participants. The grey shaded region indicates the time window 120-280 ms, when the negative going peak is identified. There seems to be a clear difference in N200 amplitude between preexposure and postexposure only for NC. ${}^{*}p < 0.05$   
		}
		\label{EPN}
	\end{center}   
\end{figure}

N200 peak was observed around 200 ms after stimulus onset for all control participants (see Figure \ref{EPN}). A significant main effect of the time factor (i.e, preexposure vs postexposure) was observed only for the NC participants (NC: $F(1,11)=5.3241, p= 0.0415$, PC: $F(1,11)=0.6882, p= 0.4244$ and NEC: $F(1,10)=1.2653, p= 0.2869$) and the exposed food items for NC at postexposure (M$= -3.88  \mu V$, SD$= 2.54$) led to significantly higher negative amplitudes ($t_{11}=2.3047, p=0.0207$) than the preexposure exposed food items (M$= -2.86  \mu V$, SD$= 3.15$). We have also observed a localization effect (left vs right) at both preexposure and postexposure only for the NEC group (see Table \ref{localization_N200} for details).  

\subsubsection*{P300}
\begin{figure}[!h]
	\begin{center}
		\includegraphics[width=18cm,height=13cm]{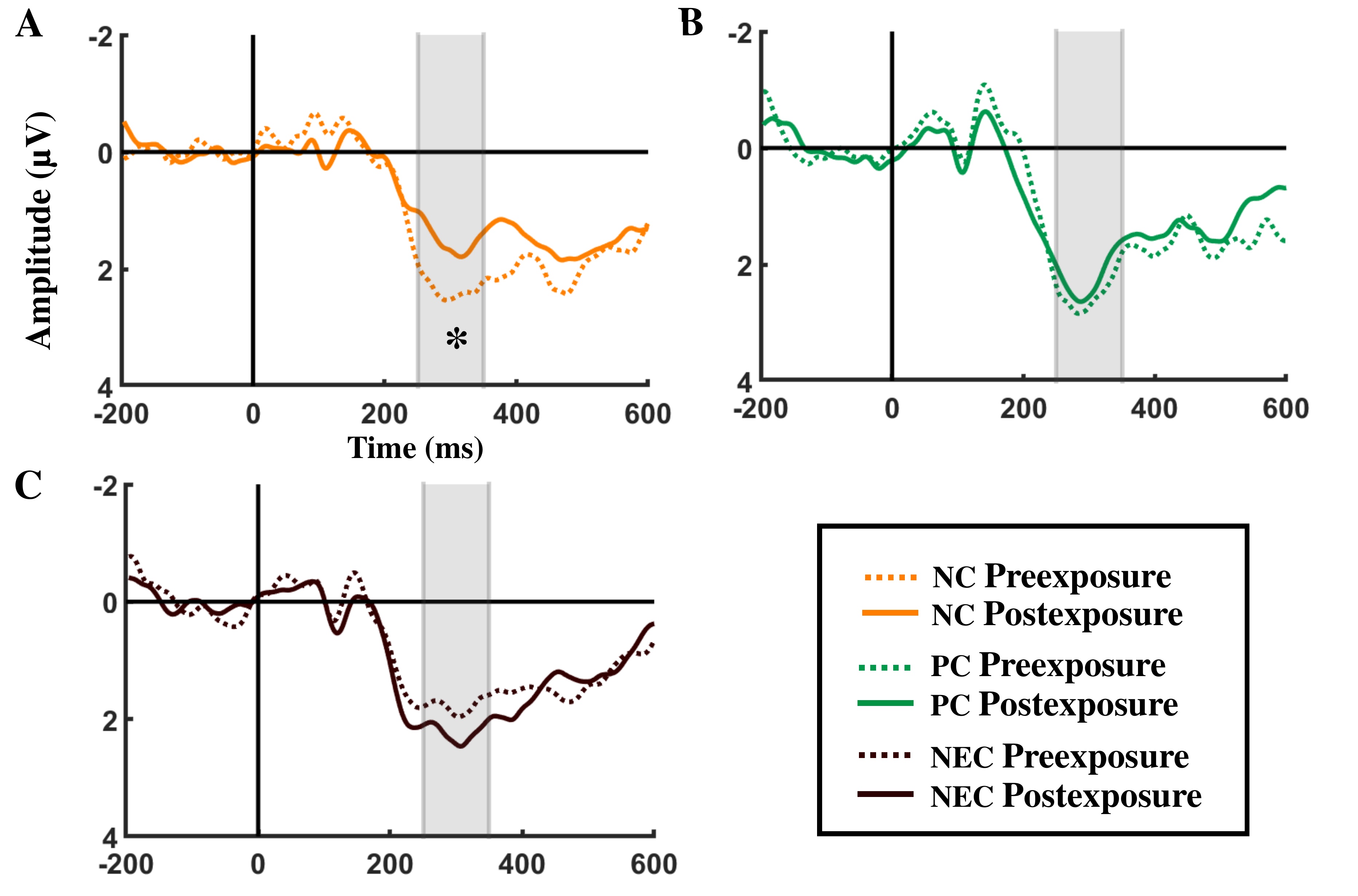}
		\caption{{\bf Event-related potentials in response to the food pictures (exposed) at the centro-parietal cluster (CP1, CP2, CP3, CP4, CPz, P1 and P2) for both preexposure and postexposure conditions.} The graph depicts the electrophysiological activity from 200 ms before to 600 ms after stimulus onset. A. ERPs for NC participants. B. ERPs for PC participants C.  ERPs for NEC participants. The grey shaded region indicates the time window, when the positive going peak is identified. There seems to be a clear difference in P300 amplitude between preexposure and postexposure only for NC. ${}^{*}p < 0.05$   
		}
		\label{LPP_P300}
	\end{center}   
\end{figure}

We have observed a positive going peak at the centro-parietal electrode cluster around 300 ms after stimulus onset for all control groups (see Figure \ref{LPP_P300} ). Among all the control groups, we have found a significant main effect of the time factor (i.e, preexposure vs postexposure) for the NC participants only (NC: $F(1,11)=5.07314,\, p= 0.0457$, PC: $F(1,11)=0.1728, \, p= 0.6856$ and NEC: $F(1,10)=0.0524, \, p= 0.8236$) indicating a more positive amplitude ($t_{11}=2.2523, \, p=0.0229$) in response to exposed food pictures at preexposure condition (M$= 3.60  \mu V$, SD$= 2.68$) as compared to the exposed food pictures at postexposure condition (M$= 2.76  \mu V$, SD$= 2.66$). We have also observed a localization effect (left vs right) at both preexposure and postexposure for the NEC group only (see Table \ref{localization_P300} for details).

\subsubsection*{Late Positive Potential}
A sustained positive amplitude from 300 ms to 600 ms at post-stimulus onset is observed for all control groups (see Figure \ref{LPP_P300}). However, there was no significant main effect of time factor (i.e, preexposure vs postexposure) for any of the control groups (NC: $F(1,11)=0.0958, p= 0.7626$, PC: $F(1,11)=0.4744, p= 0.5052$ and NEC: $F(1,10)=0.0182, p= 0.8951$). Further, We have found a localization effect (left vs right) at postexposure condition for PC and NEC (see Table \ref{localization_LPP} for details).
\subsection*{Correlation Analyses}
Partial correlation matrices of subjective desire ratings and electrophysiological
dependent variables are computed for all controls (see Table \ref{correlation} for details). NC and PC participants show moderate to very high correlations between desire and ERP components except in P200, in both preexposure and postexposure conditions. However, postexposure condition produces stronger correlations as compared to preexposure for both NC and PC. We found low correlation in case of NEC desire and ERPs.

\subsection*{Single-Trial Multivariate Analysis}
\begin{figure}[!h]
	\begin{center}
		\includegraphics[width=16cm,height=9cm]{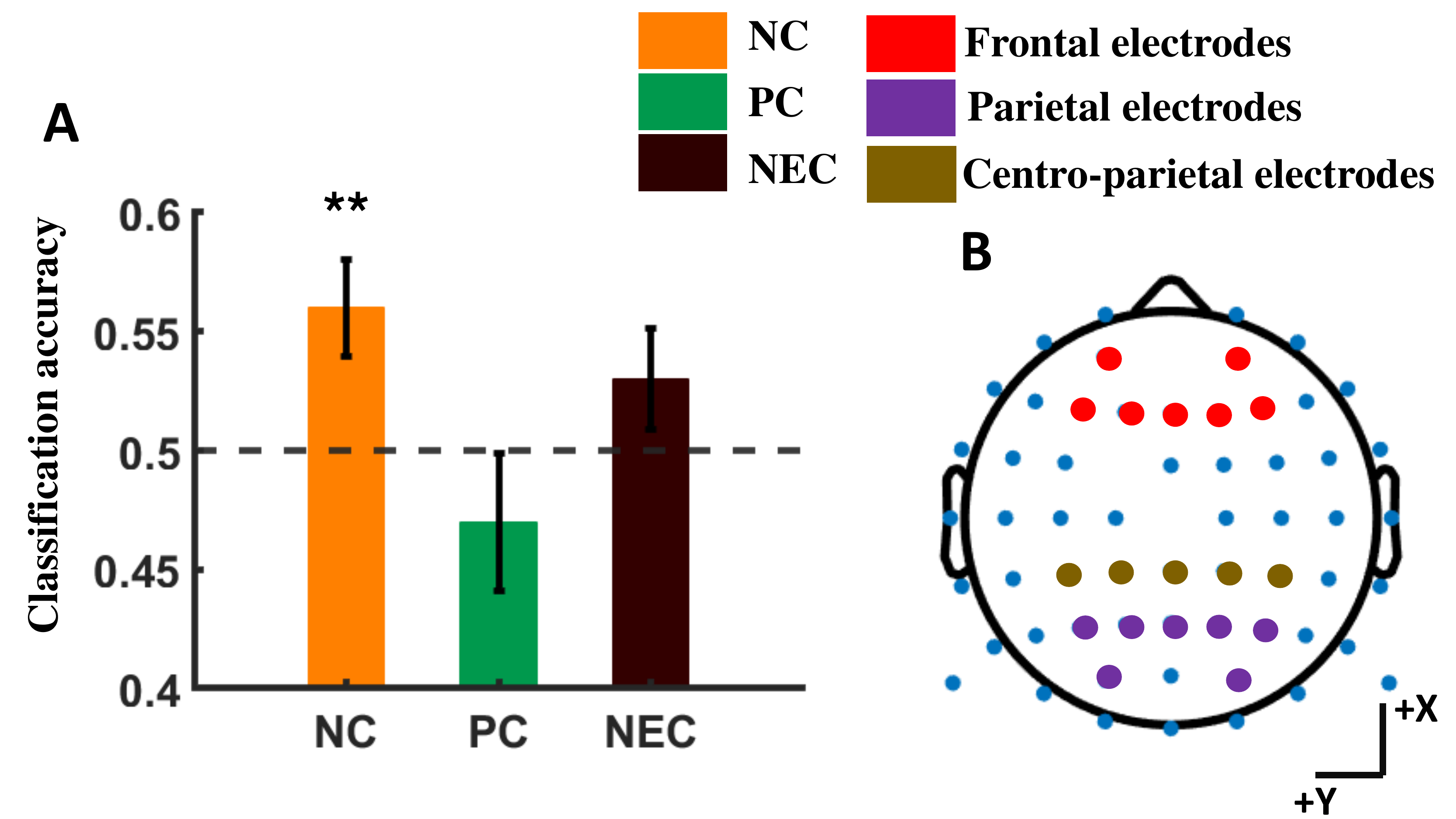}
		\caption{{\bf Classification accuracy between preexposure and postexposure for all control participants.} A. The orange, green and brown bars represent mean classification accuracies for NC, PC and NEC respectively.The horizontal dashed line represents the chance classification. B. Frontal (F1,  F2,  F3,  F4, Fz, AF3,  AF4), centro-parietal (CP1, CP2, CP3, CP4, CPz) and parietal (P1, P2, P3, P4, Pz, PO3, PO4) electrodes on a 64-channel EEG cap. There seems to be an above chance classification accuracy for the NC only. ${}^{*}p < 0.05,{}^{**}p < 0.01, {}^{***}p < 0.001$
		}
		\label{classification}
	\end{center}   
\end{figure}

A pattern classifier (CPCA) was used to analyze single-trial EEG signals corresponding to the preexposure and postexposure conditions for all controls. Only NC participants show above chance classification accuracy (NC: $0.56$, $t_{11}=3.0913, p= 0.0051$, PC: $0.47$, $t_{11}=-0.8804, p= 0.80$, NEC: $0.53$, $t_{10}=1.4072, p= 0.09$, Figure \ref{classification} ). 

 \section*{Discussion}
Rising obesity and craving related disorders including binge eating disorder necessitate detailed understanding of the neuro-cognitive mechanism behind food craving. Despite some recent work \cite{konova2018computational,boswell2018training},
our understanding of food craving remains limited. In this study, we explored craving through computational modeling and statistical analysis by subjecting participants to food cues and multi-sensory food exposure. We grouped the participants based on their choice of liked food (sweet or savory) and systematically exposed them to their liked or disliked food item.  They were grouped as positive, negative or neutral depending on their liking for exposed food items. We explored the effect of multi-sensory exposure on liked and disliked food items and computed a statistical model to predict the subjective value the participants were willing to pay to obtain the craved item. Using univariate and multivariate neural data processing, we studied the neural markers and timeline of food craving. 

\subsection*{Inducing Craving for Disliked Food Items}

Accessibility of inexpensive unhealthy food combined with aggressive marketing has led to substantial rise in craving related disorders in the last decade. Thus there is an urgent need to  develop efficient intervention strategies to promote healthy eating. Public policy interventions include strategies like reducing availability or increasing price of unhealthy food \cite{liu2014using}.
Intervention studies have tried to address this issue by using behavioral \cite{liu2014using, schwab2017disgust},
financial \cite{harris2013temporally} and cognitive strategies \cite{kober2010regulation,svaldi2015effects,meule2013time,sarlo2013cognitive,boswell2018training}.
One approach in these intervention studies includes nudging people to healthier food choices mostly using priming and saliency \cite{wilson2016nudging}, 
although the efficacy of such studies is mixed. Most of the intervention studies focus on down regulating craving using cognitive strategies and had consistently shown that delaying gratification \cite{mischel1975cognitive,meule2013time}, 
thinking about negative long term consequences reduce craving for both food \cite{giuliani2014neural,boswell2018training} 
and drugs \cite{boswell2018training,kober2010regulation}.
Our results show that it is possible to up regulate food craving for previously disliked food items by exposing participants to disliked food. By grouping participants based on their liking of particular food items, we were able to systematically observe the effect of food cues and food exposure on liked and disliked food items.  Our results establish that multi-sensory food exposure was instrumental in inducing craving irrespective of likeness for particular food item. Thus, for participants disliking a particular food item (Negative control participants), it is possible to induce craving of disliked food by subjecting them to multi-sensory food exposure. Our results also show significant reduction in desire values for previously liked food items for negative control participants following exposure to disliked food thus showing a potential reversal of craving preference. Craving effect for disliked food also translated to subjective values of food item and willingness to pay and it could potentially contribute to better intervention studies and have future policy implications. 
Although in our study, both disliked and liked food items consist of high-calorie food, but follow up studies using similar paradigms consisting of both high calorie and low- calorie food could be pursued. 

\subsection*{Role of Food Exposure and Willingness to Pay in Craving }

Food craving studies generally show that desire for exposed or displayed food items increases the momentarily desire and subjective value of the food item.  Majority of studies exploring food choices and craving use food cues and mental imagery in the experimental paradigm and to the best of our knowledge, only a couple of studies exist \cite{konova2018computational,wolz2017subjective}  where multi-sensory exposure was included in the experimental paradigm. In this study, we demonstrated that multi-sensory exposure to food items produces significant effect on craving as compared to food cues. Our behavioral analyses showed that post exposure desire and bidding ratings were always higher for the food item and this effect is significant in negative and neutral control participants (see Table \ref{table1} and Figure \ref{fig2}). 
FCQ-S scores pre and post experiment showed that craving was induced for all three categories of participants (positive, negative and neutral) significantly post experiment (see Figure \ref{fig3}). Our results also showed that there is no time dependency between inducement of craving and last meal taken by participants (see Table \ref{time_duration}). Hence for craving related research, it is not necessary to bar participants from having food for a long time, a fact often ignored by most studies \cite{konova2018computational,lowe2018neurocognitive}. 

In majority of the studies, the participant's response is typically given in terms of inconsequential desire rating for the displayed food cues. However, whether desire for a particular food item translates to WTP often remains unexplored \cite{konova2018computational}. Our results show that increase in desire does not always lead to WTP.
In case of neutral control participants, multiple analyses showed that although desire for exposed food items increased momentarily, it did not translate to significant increase in concurrent subjective valuation of food item which was measured by the bidding scale when compared with pre-exposure bidding data. This is an important result from the perspective of consumer behavior and establishes the importance of recording subjective valuation of food cues.

\subsection*{Neural Mechanism of Craving}
Cognitive functioning plays a role in modulating food craving and can help elucidate the underlying neural mechanism. Neuroimaging studies have revealed that food cues inducing craving leads to significant neural activity in ventral striatum (VS) \cite{kober2010prefrontal,o2006predictive}, 
amygdala \cite{due2002activation,franklin2007limbic}, insula \cite{franklin2007limbic,wilson2005instructed}, ventral tegmental area (VTA) \cite{kober2010prefrontal,due2002activation}, orbitofrontal cortex \cite{kober2010prefrontal,due2002activation,david2005ventral,franklin2007limbic} and anterior cingulate cortex (ACC) \cite{kober2010prefrontal,due2002activation,david2005ventral}. Most of these regions are known to modulate emotion, motivation and are part of dopamine reward pathway \cite{di1988drugs,haber2010reward}. Similar results are also found in drug addiction studies \cite{pelchat2004images,tang2012food,schienle2009binge}.  

The neural timeline modulating food craving could be reflected by ERP components measured during the experimental paradigm. Initial sensory and attention allocation towards food cues are reflected in increased P200, N200 components \cite{carbine2018utility,franken2011electrophysiology} as compared to neutral stimuli. Higher-order attentional allocation was reflected in increased P300 and LPP for more palatable food \cite{sarlo2013cognitive}, neutral stimuli \cite{kong2015inhibition} or less emotionally charged food \cite{stockburger2009vegetarianism,versace2016heterogeneity}.

We have shown using behavioral analyses and modeling that participants disliking particular food item (negative control participants) on exposure to the food reverses her/his previous preference and develops craving for exposed food. Our behavioral results are also corroborated using multivariate pattern analysis on neural signals showing a significant difference between pre and post exposure for negative control participants. Unlike most previous studies, in the current setup, participants are exposed to food cues throughout the experiment and we compare the ERP amplitudes pre and post multi-sensory exposure of food. Our results show that there is no significant difference between participant categories for P200 and LPP whereas for the negative control participants, N200 and P300 are significantly different pre and post food exposure. An increase in parietal N200 and decrease in centroparietal P300 post exposure of disliked food was observed for negative control participants. 

We hypothesize that increase in parietal N200 and a decrease in P300 following food exposure possibly act as neural markers for producing increased desire for previously disliked food items. Increase in N200 has been previously associated with high calorie food \cite{kong2015inhibition},  initial  sensory or visual attention (see \cite{carbine2018utility} for review). Increase in craving in negative control participants following food exposure is seen to be associated with increase in N200, possibly signifying an increase in initial allocation of attention and decreased inhibition to a previously disliked food item.

It is interesting to note that we get significant difference in the P300 components which is typically found in food-related inhibitory control and intervention studies \cite{luijten2014systematic,patrick2006p300,kamarajan2010dysfunctional}. In intervention studies, decrease in P300 has been related to cognitive regulation strategies \cite{svaldi2015effects} and short-term consequences for consuming high-calorie food \cite{meule2013time}. P300 is a known ERP marker for response inhibition \cite{hachl2003erps} and while frontal N200 can reflect initial inhibition processing, P300 can account for increased recruitment for inhibitory control processes \cite{wessel2015s,albert2013spatiotemporal}. A recent craving study advocates the role of only P300 and not N200 in inhibitory control deficits in obese people. In the study by Kong et al. \cite{kong2015inhibition}, an increase in P300 for neutral and low-calorie food compared to high-calorie food for successful restrained eaters was shown pointing to the role of larger P300 amplitude for increased recruitment of inhibitory control process. In the current setup, pre-exposure conditions for negative control participants produce similar scenario as seen in inhibitory control studies leading to an increase in P300 component for disliked food. However, following food exposure, the preference for exposed food for negative control participants is seen to be reversed which is possibly reflected in lowering of inhibitory control and hence associated with decrease in P300 amplitude. 


\section*{Conclusion}
The current study aims at manipulating food craving preferences using multi-sensory food exposure. Grouping participants based on their food preferences, the study shows that craving can be induced through food exposure even for disliked food items. Neural markers modulating craving are explored and the results seem to point to the role of reduction of inhibitory control for reversing momentary craving preference. Further studies in modulating food craving following similar paradigms could potentially have a long-term impact on public health policy.

\section*{Funding }
A. Chatterjee is supported by an INSPIRE fellowship (no: IF170367) from the Department of Science and Technology (DST), Government of India.
\section*{Declarations of interest }
None.
\section*{Author contributions }
A.C. and K.D. designed the research. A.C. and S.M. performed the research. A.C., S.M. and K.D. wrote the manuscript. K.D.
edited the manuscript and supervised the entire work. All authors have read and approved the final manuscript.
\section*{Data availability}
The data that support the findings of this study are available from the corresponding authors upon reasonable request. 
\section*{Ethical statement}
Written informed consent was obtained from each participant before
the experiment. The experimental protocol was in accordance with the
Declaration of Helsinki and approved by the the Institute Ethics Committee of IISER, Kolkata. 

\newcounter{NoTableEntry}
\renewcommand*{\theNoTableEntry}{NTE-\the\value{NoTableEntry}}
\makeatletter
\newcommand*{\notableentry}{%
  \kern-\tabcolsep
  \stepcounter{NoTableEntry}%
  \vadjust pre{\zsavepos{\theNoTableEntry t}}
  \vadjust{\zsavepos{\theNoTableEntry b}}
  \zsavepos{\theNoTableEntry l}
  \raisebox{%
    \dimexpr\zposy{\theNoTableEntry b}sp
    -\zposy{\theNoTableEntry l}sp\relax
  }[0pt][0pt]{%
    \color{black}%
    \setlength{\unitlength}{1pt}%
    \edef\w{%
      \strip@pt\dimexpr\zposx{\theNoTableEntry r}sp%
      -\zposx{\theNoTableEntry l}sp\relax
    }%
    \edef\h{%
      \strip@pt\dimexpr\zposy{\theNoTableEntry t}sp%
      -\zposy{\theNoTableEntry b}sp\relax
    }%
    \ifdim\w pt=0pt 
    \else
      \begin{picture}(0,0)%
        \edef\x{%
          \noexpand\put(0,0){\noexpand\line(\w,\h){\w}}%
          \noexpand\put(0,\h){\noexpand\line(\w,-\h){\w}}%
        }\x
      \end{picture}%
    \fi
  }%
  \hspace{0pt plus 1filll}%
  \zsavepos{\theNoTableEntry r}
  \kern-\tabcolsep
}   

\setcounter{figure}{0}
\makeatletter 
\renewcommand{\thefigure}{S\@arabic\c@figure}
\newcounter{SIfig}
\renewcommand{\theSIfig}{S\arabic{SIfig}}
\makeatother

\setcounter{table}{0}
\makeatletter 
\renewcommand{\thetable}{S\@arabic\c@table}
\newcounter{SItab}
\renewcommand{\theSItab}{S\arabic{SItab}}
\makeatother
\begingroup
    \fontsize{9pt}{10pt}\selectfont
    
\section*{Supporting Information} 

\subsection*{PC, NEC and NC class formation}
At first, we have calculated the difference ($\Delta_i$) between the mean value ratings of chocolate and chips for all subjects $i$. Then we put $i$ th subject in NEC if $\Delta_i$ lies between the interval ($-\frac{s}{\sqrt{n}}t_{\mbox{\tiny{(n-1)}},\frac{\alpha}{2}}$, $\frac{s}{\sqrt{n}}t_{\mbox{\tiny{(n-1)}},\frac{\alpha}{2}}$) else in the NC or PC class, where $n=96$ (number of participants), $s=$ standard deviation of $\Delta$ (which was found to be $2.72$) and $t_{\mbox{\tiny{(n-1)}},\frac{\alpha}{2}}$ is the $(1-\frac{\alpha}{2})$th quantile of t-distribution with $(n-1)$ degrees of freedom.  In our analysis we have chosen $\alpha$ = 0.05.

\begin{table}[!h]
\begin{center}
\setlength{\arrayrulewidth}{0.3mm}
\begin{tabular}{lccc}
\hline
\multicolumn{1}{l}{}&\multicolumn{1}{c}{Shorter duration}&\multicolumn{1}{c}{Longer duration}&\multicolumn{1}{c}{Test Statistic}\tabularnewline
&\multicolumn{1}{c}{{\scriptsize $N=18$}}&\multicolumn{1}{c}{{\scriptsize $N=17$}}&\tabularnewline
\hline
~~~~YES craving&72\%~{\scriptsize~(13)}&65\%~{\scriptsize~(11)}&$ \chi^{2}(1)=0.2292 ,~ P=0.6321  $\tabularnewline
~~~~NO craving&28\%~{\scriptsize~(5)}&35\%~{\scriptsize~(6)}&\tabularnewline
\hline
\end{tabular}
\refstepcounter{SItab}\label{time_duration_EEG}
\caption{Time duration difference in response to question of whether food cravings had ever been
experienced. This analysis is done based on the behaviourall data of the participants who took part in EEG experiment.}
\end{center}

\end{table}

\subsection*{Time duration of refrained from eating}
The chi-square (${\chi}^2$) test of independence  is performed to test the dependency between time duration of refrained from eating and the inducement of craving on the subjects performing the EEG experiment ($N=35$) based on a $2 \times 2$ contingency table (see Table \ref{time_duration_EEG}). Two classes (viz., Shorter duration and Longer duration) were formed based on the median time duration (found to be 193 min). Based on the median value, two classes viz., Shorter duration (120 min $\leq$ time $\leq$ 193 min) and Longer duration ($>$ 193 min) were formed. A participant is classified in the Shorter duration class (or Longer duration class) if the participant's time duration of refrained from eating  belongs to (120, 193] (or (193, 360]).
The test confirmed that there is no reason to believe that food craving is dependent on the time duration of refrained from eating ($\chi^{2}(1)=0.2292$, p-value $= 0.6321$, see Table \ref{time_duration_EEG}).

\subsection*{Spline Regression Model (SRM)}
A $k$th degree spline regression model with h knots, $t_1<t_2<...<t_h$, with no continuity restriction is given by:
\begin{equation}
  y=\alpha +\sum_{j=1}^{h+1}\sum_{i=0}^{k}\beta_{ij}C_j(x)x^i+\epsilon,  
\end{equation}
where $\alpha$ and $\beta_{ij}$ are the coefficients of the model and $\epsilon$ follows $N(0,\sigma^2)$. $C_{j}(x)$ is defined as follows:
\begin{align*}
    C_{0}(x) &= I(x< t_1) \\
    C_{1}(x) &= I(t_1\leq x < t_2) \\
    &~~ \vdots \\
    C_{h-1}(x) & = I(t_{h-1}\leq x < t_{h})\\
    C_{h-1}(x) & = I(t_{h}\leq x)
\end{align*}
\subsection*{Bonus amount calculation}
Becker-Degrot-Marschak (BDM) is a commonly used auction procedure to elicit  willingness-to-pay. In general, under the BDM, a participant announces a bid for an item; the item's price is then randomly drawn. If the bid amount exceeds the price, the participant gets the item and pays the drawn price. If the bid amount is less than the drawn price, the participant does not receive the item and pays nothing. 
  In each trial of our study, the proportion of the bid amount out of the left amount was calculated. Then a random number between 0 and 1 was generated. If this proportion exceeds the random number, the participants won an extra unit of rupee~10 and if the proportion is less than the random number, the participant gets no extra unit but the bid amount was deducted from the total endowment given to the block unlike the standard BDM. Then a random block was selected for each participant and total number of the won  unit was counted for that block. A subject receives rupee~100 + number of won unit$\times$ rupee~10.\\
  Bonus amount was calculated in the similar fashion for the EEG studies. But each of the subjects for the EEG study was compensated at the consolidated rate of Rs. rupee~400 + number of won unit$\times$ rupee~10.
  \newpage
\subsection*{Tables}
\begin{table}[h]
\centering
\begin{tabular}{llllll}
Source       & Sum   Sq. & d.f. & Mean Sq. & F     & Prob\textgreater{}F \\\hline
Time         & 16.327    & 1    & 16.3274  & 23.71 & 0.0000            \\
Control      & 9.282     & 2    & 4.6412   & 6.74  &0.0015            \\
Time*Control & 9.302     & 2    & 4.6512   & 6.75  & 0.0015           \\
Error        & 130.166   & 189  & 0.6887   &       &                  \\
Total        & 186.76    & 194  &          &       &                     \\\hline
\end{tabular}
\refstepcounter{SItab}\label{ANOVA1}
\caption{\bf {ANOVA table for exposed bid series}}
\end{table}

\begin{table}[h]
\centering
\begin{tabular}{llllll}
Source       & Sum   Sq. & d.f. & Mean Sq. & F     & Prob\textgreater{}F \\\hline
Time         & 22.894    & 1    & 22.8941  & 59.87 & 5.92$\times 10^{-13}$  \\
Control      & 28.953    & 2    & 14.4765  & 37.85 & 1.49$\times 10^{-14}$  \\
Time*Control & 10.73     & 2    & 5.3649   & 14.03 & 2.09$\times 10^{-06}$  \\
Error        & 72.278    & 189  & 0.3824   &       &                     \\
Total        & 146.979   & 194  &          &       &                      \\\hline

\end{tabular}
\refstepcounter{SItab}\label{ANOVA2}
 \caption{\bf {ANOVA table for exposed desire series}}
\end{table}


\begin{table}[!ht]
\centering
\begin{tabular}{lllllll|lllll} 
\toprule
\multicolumn{12}{l}{~ ~ ~ ~ ~ ~ ~ ~ ~ ~ ~ ~ ~ ~ ~ ~ ~  \textbf{Preexposure}~ ~ ~ ~ ~ ~ ~ ~ ~ ~ ~ ~ ~ ~ ~ ~ ~\textbf{Postexposure}}             \\\hline
    &        & Desire   & P200     & N200      & P300     & LPP      & Desire   & P200     & N200      & P300     & LPP       \\
    & Desire & \textit{\textbf{1}}        & \textbf{0.30} & \textbf{0.30} & -0.18 & 0.11 & \textit{\textbf{1}}        &\textit{\textbf{0.59}}  & \textit{\textbf{0.58}} & \textit{\textbf{-0.46}} &\textbf{0.39}   \\
    & P200   & \textbf{0.30} & \textit{\textbf{1}}        & \textit{\textbf{-0.92}}  & \textit{\textbf{0.44}}  & -0.21 &\textit{\textbf{0.59}}   & \textit{\textbf{1}}        & \textit{\textbf{-0.96}}  & \textit{\textbf{0.80}}  & \textit{\textbf{-0.73}}   \\
NC  & N200    & \textbf{0.30} &\textit{\textbf{-0.92}}   & \textit{\textbf{1}}         & \textit{\textbf{0.56}}  & -0.18 & \textit{\textbf{0.58}} &\textit{\textbf{-0.96}}   & \textit{\textbf{1}}        &\textit{\textbf{0.81}}   & \textit{\textbf{-0.69}}  \\
    & P300   & -0.18 & \textit{\textbf{0.44}}   & \textit{\textbf{0.56}}  & \textit{\textbf{1}}         &\textit{\textbf{0.68}}  & \textit{\textbf{-0.46}}  &\textit{\textbf{0.80}}   & \textit{\textbf{0.81}}  & \textit{\textbf{1}}        & \textit{\textbf{0.93}} \\
    & LPP    & 0.11 & -0.21 & -0.18 & \textit{\textbf{0.68}} & \textit{\textbf{1}}        & \textbf{0.39} & \textit{\textbf{-0.73}} & \textit{\textbf{-0.69}}  & \textit{\textbf{0.93}} & \textit{\textbf{1}}         \\\hline
    & Desire & \textit{\textbf{1}}        & -0.05 & -0.27 & -0.24 & \textbf{0.38} & \textit{\textbf{1}}        & 0.28 & 0.26 & -0.35 & \textit{\textbf{0.43}}  \\
    & P200   & -0.05 & \textit{\textbf{1}}        & \textit{\textbf{-0.81}} & -0.08 & 0.13 & 0.28 & \textit{\textbf{1}}        & \textit{\textbf{-0.89}}  & 0.12 & -0.12  \\
PC  & N200    & -0.27 & \textit{\textbf{-0.81}} & \textit{\textbf{1}}        & -0.13 & 0.23 & 0.26& \textit{\textbf{-0.89}}  & \textit{\textbf{1}}        & 0.21 & -0.29  \\
    & P300   & -0.24 & -0.08 & -0.13 & \textit{\textbf{1}}        & \textit{\textbf{0.72}}  & \textbf{-0.35}& 0.12 & 0.21 & \textit{\textbf{1}}        & \textit{\textbf{0.73}}   \\
    & LPP    & \textbf{0.38} & 0.13 & 0.23 & \textit{\textbf{0.72}} & \textit{\textbf{1}}        &\textit{\textbf{0.43}} & -0.12& -0.29 &\textit{\textbf{0.73}}  & \textit{\textbf{1}}         \\\hline
    & Desire & \textit{\textbf{1}}        & -0.001 & -0.25 & -0.27& 0.06  & \textit{\textbf{1}}        & 0.05 & -0.09 & -0.08& -0.21 \\
    & P200   & -0.001 & \textit{\textbf{1}}        & \textit{\textbf{-0.89}}& -0.27 & 0.06 & 0.05 & \textit{\textbf{1}}        & \textit{\textbf{-0.94}}   & 0.06  & -0.23 \\
NEC & N200    & -0.25& \textit{\textbf{-0.89}}& \textit{\textbf{1}}        & \textit{\textbf{-0.42}} & 0.21 & -0.09 & \textit{\textbf{-0.94}}   & \textit{\textbf{1}}        & -0.02 & -0.22  \\
    & P300   & -0.27 & -0.27 & \textit{\textbf{-0.42}} & \textit{\textbf{1}}        &\textit{\textbf{0.89}}  & -0.08 & 0.06  & -0.02 & \textit{\textbf{1}}        & \textit{\textbf{0.72 }}  \\
    & LPP    & 0.06  & 0.06 & 0.21 & \textit{\textbf{0.86}}  & \textit{\textbf{1}}        & -0.21 & -0.23 & -0.22 & \textit{\textbf{0.72}}  & \textit{\textbf{1}}         \\
\bottomrule

\end{tabular}

\refstepcounter{SItab}\label{correlation}

\caption{Matrix of partial correlations for ERP amplitudes with desire rating at preexposure and postexposure condition. High correlations (i.e, $\lvert r \rvert \geq 0.3$ ) are indicated in bold. Very high correlations (i.e, $\lvert r \rvert \geq 0.4$  ) are indicated in bold italic. }
\end{table}
\begin{table}
  \centering
  \renewcommand{\arraystretch}{1.2}
  \setlength{\arrayrulewidth}{.3mm}
  \rowcolors{4}{}{}
  \begin{tabular}{|p{1.2cm}|c|c|c|c|c|}
  
    \hline
    \multirow{2}{5cm}{\textbf{}} & \multicolumn{2}{c|}{\textbf{Preexposure}} & \multicolumn{2}{c|}{\textbf{Postexposure}}\\
    \cline{2-5}
    & \cellcolor{gray!60!yellow!70!}\textbf{Mean difference, t-value} &\cellcolor{gray!60!yellow!70!} \textbf{$p-$value} & \cellcolor{gray!60!yellow!70!}\textbf{Mean difference, t-value} &\cellcolor{gray!60!yellow!70!} \textbf{$p-$value}\\
    \hline
   \bf NC & $-1.8634, t_{11}=-1.6176$ & $0.0670$ & $-0.6241,t_{11}=-1.0130$ & $0.1664$ \\ \hline
    \bf  PC & $-1.0627,t_{11}=-0.9043$ & $0.1926$ &$0.0081,t_{11}=0.0241$ & $0.4906$ \\ \hline
  \bf NEC & $-0.0071,t_{10}=-0.0116$ & $0.4955$ &$0.3313,t_{10}=1.01331$ & $0.1674$  \\ \hline
    
  \end{tabular}
  \caption{\textbf{Localization effect of P200.} Paired t-test of P200 amplitudes between left (F1, F3, AF3) vs right (F2, F4, AF4) electrode clusters at preexposure and postexposure are performed and corresponding t- value, mean difference (right-left) and $p-$ value are mentioned.}
  \label{localization_P200}
\end{table}

\begin{table}
  \centering
  \renewcommand{\arraystretch}{1.2}
  \setlength{\arrayrulewidth}{.3mm}
  \rowcolors{4}{}{}
    \begin{tabular}{|p{1.2cm}|c|c|c|c|c|}
  
    \hline
    \multirow{2}{5cm}{\textbf{}} & \multicolumn{2}{c|}{\textbf{Preexposure}} & \multicolumn{2}{c|}{\textbf{Postexposure}}\\
    \cline{2-5}
    & \cellcolor{gray!60!yellow!70!}\textbf{Mean difference, t-value} &\cellcolor{gray!60!yellow!70!} \textbf{$p-$value} & \cellcolor{gray!60!yellow!70!}\textbf{Mean difference, t-value} &\cellcolor{gray!60!yellow!70!} \textbf{$p-$value}\\
    \hline
   \bf NC & $0.1507, t_{11}=0.1667$ & $0.4353$ & $0.4043,t_{11}=0.6020$ & $0.2797$ \\ \hline
    \bf  PC & $-2.1902,t_{11}=-1.3996$ & $0.0946$ &$-1.0275,t_{11}=-0.7005$ & $0.2491$ \\ \hline
  \bf NEC & $1.8796,t_{10}=3.5183$ & $0.0028$** &$1.1236,t_{10}=2.2424$ & $0.0244$*  \\ \hline
    
  \end{tabular}
  \caption{\textbf{Localization effect of N200.} Paired t-test of N200 amplitudes between left (P1, P3, PO3) vs right (P2, P4, PO4) electrode clusters at preexposure and postexposure are performed and corresponding t- value, mean difference (right-left) and $p-$ value are mentioned. ${}^{*}p < 0.05,{}^{**}p < 0.01, {}^{***}p < 0.001$}
  \label{localization_N200}
\end{table}

\begin{table}
  \centering
  \renewcommand{\arraystretch}{1.2}
  \setlength{\arrayrulewidth}{.3mm}
  \rowcolors{4}{}{}
    \begin{tabular}{|p{1.2cm}|c|c|c|c|c|}
  
    \hline
    \multirow{2}{5cm}{\textbf{}} & \multicolumn{2}{c|}{\textbf{Preexposure}} & \multicolumn{2}{c|}{\textbf{Postexposure}}\\
    \cline{2-5}
    & \cellcolor{gray!60!yellow!70!}\textbf{Mean difference, t-value} &\cellcolor{gray!60!yellow!70!} \textbf{$p-$value} & \cellcolor{gray!60!yellow!70!}\textbf{Mean difference, t-value} &\cellcolor{gray!60!yellow!70!} \textbf{$p-$value}\\
    \hline
   \bf NC & $-0.0844, t_{11}=-0.1053$ & $0.4590$ & $0.4954,t_{11}=1.0829$ & $0.1510$ \\ \hline
    \bf  PC & $0.4403,t_{11}=0.5765$ & $0.2879$ &$0.6905,t_{11}=0.8261$ & $0.2132$ \\ \hline
  \bf NEC & $1.1400,t_{10}=2.4101$ & $0.0183$* &$1.0570,t_{10}=2.5577$ & $0.0142$*  \\ \hline
    
  \end{tabular}
  \caption{\textbf{Localization effect of P300.} Paired t-test of P300 amplitudes between left (CP1, CP3, P1) vs right (CP2, CP4, P2) electrode clusters at preexposure and postexposure are performed and corresponding t- value, mean difference (right-left) and $p-$ value are mentioned. ${}^{*}p < 0.05,{}^{**}p < 0.01, {}^{***}p < 0.001$}
  \label{localization_P300}
\end{table}
\begin{table}
  \centering
  \renewcommand{\arraystretch}{1.2}
  \setlength{\arrayrulewidth}{.3mm}
  \rowcolors{4}{}{}
    \begin{tabular}{|p{1.2cm}|c|c|c|c|c|}
  
    \hline
    \multirow{2}{5cm}{\textbf{}} & \multicolumn{2}{c|}{\textbf{Preexposure}} & \multicolumn{2}{c|}{\textbf{Postexposure}}\\
    \cline{2-5}
    & \cellcolor{gray!60!yellow!70!}\textbf{Mean difference, t-value} &\cellcolor{gray!60!yellow!70!} \textbf{$p-$value} & \cellcolor{gray!60!yellow!70!}\textbf{Mean difference, t-value} &\cellcolor{gray!60!yellow!70!} \textbf{$p-$value}\\
    \hline
   \bf NC & $-0.0946, t_{11}=-0.1137$ & $0.4558$ & $0.0800,t_{11}=0.1748$ & $0.4322$ \\ \hline
    \bf  PC & $0.4837,t_{11}=0.5429$ & $0.2990$ &$1.2122,t_{11}=2.0609$ & $0.0319$* \\ \hline
  \bf NEC & $0.5418,t_{10}=1.3337$ & $0.1059$ &$0.7893,t_{10}=2.0424$ & $0.0342$*  \\ \hline
    
  \end{tabular}
  \caption{\textbf{Localization effect of LPP.} Paired t-test of LPP  between left (CP1, CP3, P1) vs right (CP2, CP4, P2) electrode clusters at preexposure and postexposure are performed and corresponding t- value, mean difference (right-left) and $p-$ value are mentioned. ${}^{*}p < 0.05,{}^{**}p < 0.01, {}^{***}p < 0.001$}
  \label{localization_LPP}
\end{table}

\clearpage

\subsection*{Scatter plots of NC and PC }

\begin{figure}[!ht]
\label{scatterplot}
\begin{center}
    \includegraphics[width=16cm,height=12cm]{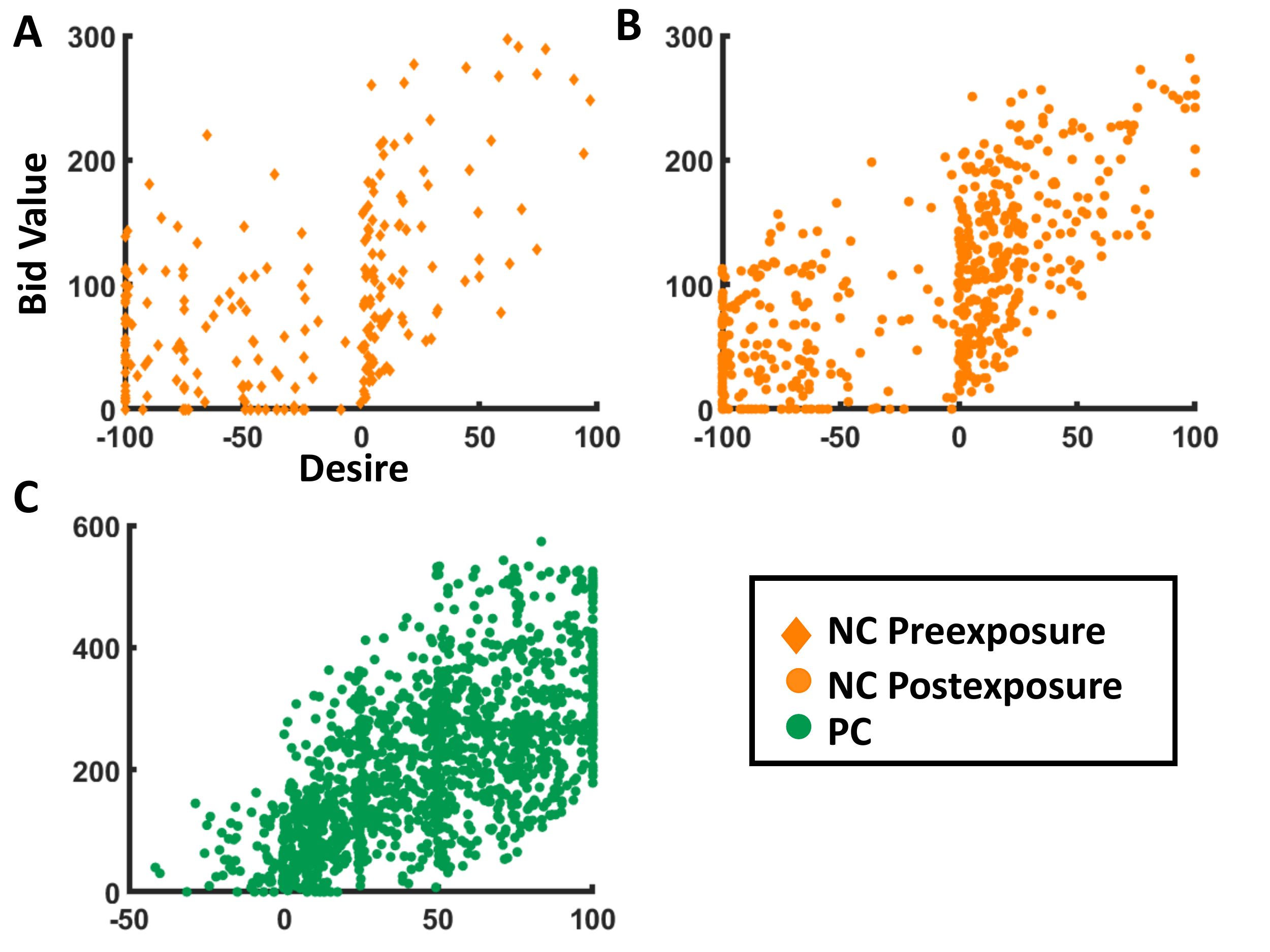}
    
    \refstepcounter{SIfig}\label{s3}
    
    \caption{\bf {Scatter plots of NC and PC.} In order to visualise the data well, bid value proportion is multiplied by 1000 A: NC at Preexposure B: NC at Postexposure C: PC }
    
\end{center}   
\end{figure}

\subsection*{Change point for nonexposed food item }

\begin{figure}[!ht]
\begin{center}
    \includegraphics[width=16cm,height=10cm]{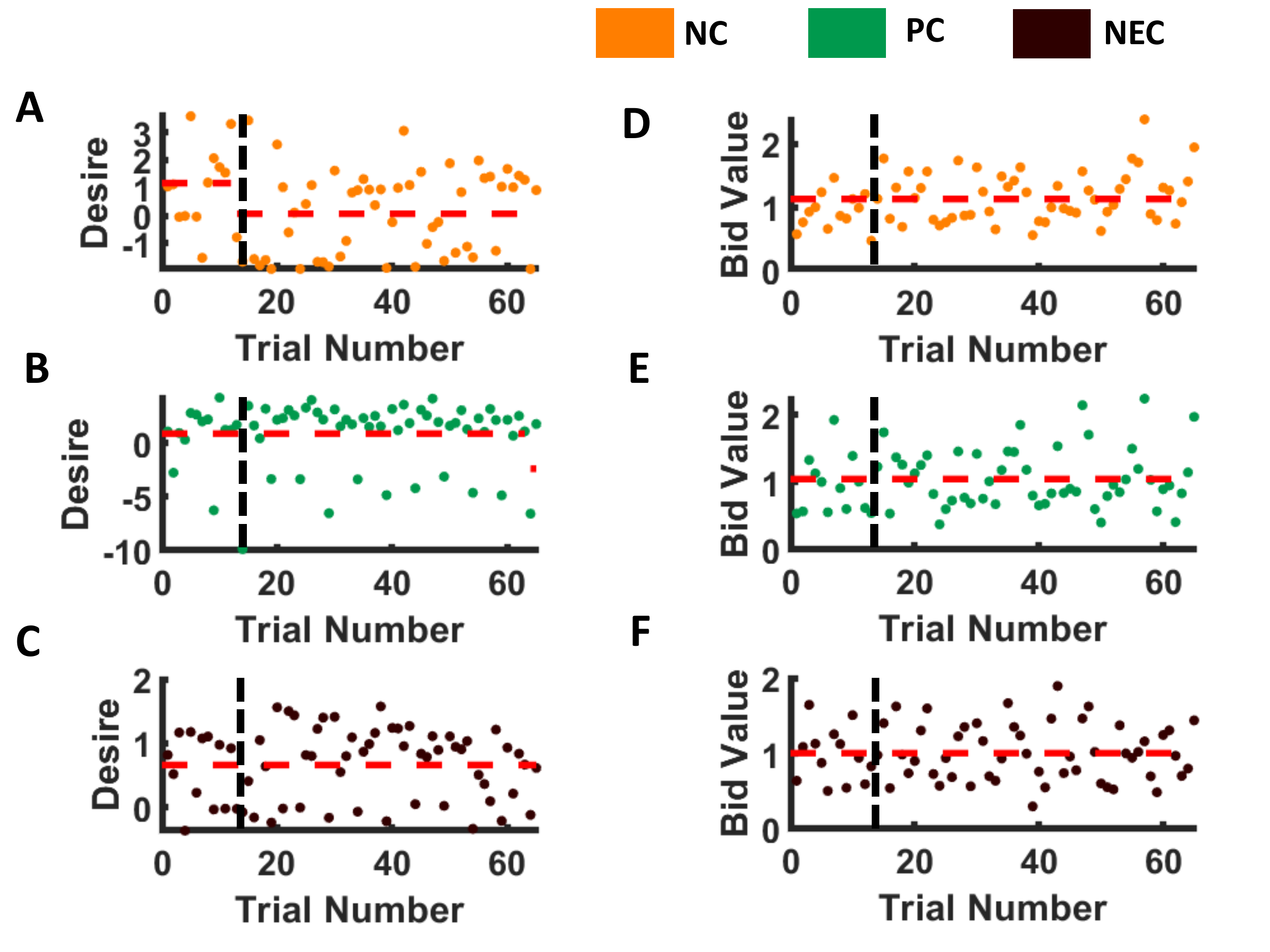}
    
    \refstepcounter{SIfig}\label{s4}
    
    \caption{{\bf Change point detection for nonexposed series.}
     Scatter plot of all controls for both bid value and desire value nonexposed foods. Vertical black dashed lines indicate trial 15, after which multisensory food exposure has occurred. Red dashed lines indicate estimated mean value and a jump in that line indicates the occurrence of a change point. } 
    
\end{center}   
\end{figure}
	
		\bibliographystyle{abbrvnat}
	\bibliography{myref1}

\begin{thebibliography}{67}
\providecommand{\natexlab}[1]{#1}
\providecommand{\url}[1]{\texttt{#1}}
\expandafter\ifx\csname urlstyle\endcsname\relax
  \providecommand{\doi}[1]{doi: #1}\else
  \providecommand{\doi}{doi: \begingroup \urlstyle{rm}\Url}\fi

\bibitem[Albert et~al.(2013)Albert, L{\'o}pez-Mart{\'\i}n, Hinojosa, and
  Carreti{\'e}]{albert2013spatiotemporal}
J.~Albert, S.~L{\'o}pez-Mart{\'\i}n, J.~A. Hinojosa, and L.~Carreti{\'e}.
\newblock Spatiotemporal characterization of response inhibition.
\newblock \emph{Neuroimage}, 76:\penalty0 272--281, 2013.

\bibitem[Allen and Seaman(2007)]{allen2007likert}
I.~E. Allen and C.~A. Seaman.
\newblock Likert scales and data analyses.
\newblock \emph{Quality progress}, 40\penalty0 (7):\penalty0 64--65, 2007.

\bibitem[Asmaro et~al.(2012)Asmaro, Jaspers-Fayer, Sramko, Taake, Carolan, and
  Liotti]{asmaro2012spatiotemporal}
D.~Asmaro, F.~Jaspers-Fayer, V.~Sramko, I.~Taake, P.~Carolan, and M.~Liotti.
\newblock Spatiotemporal dynamics of the hedonic processing of chocolate images
  in individuals with and without trait chocolate craving.
\newblock \emph{Appetite}, 58\penalty0 (3):\penalty0 790--799, 2012.

\bibitem[Becker et~al.(1964)Becker, DeGroot, and Marschak]{becker1964measuring}
G.~M. Becker, M.~H. DeGroot, and J.~Marschak.
\newblock Measuring utility by a single-response sequential method.
\newblock \emph{Behavioral science}, 9\penalty0 (3):\penalty0 226--232, 1964.

\bibitem[Boswell et~al.(2018)Boswell, Sun, Suzuki, and
  Kober]{boswell2018training}
R.~G. Boswell, W.~Sun, S.~Suzuki, and H.~Kober.
\newblock Training in cognitive strategies reduces eating and improves food
  choice.
\newblock \emph{Proceedings of the National Academy of Sciences}, 115\penalty0
  (48):\penalty0 E11238--E11247, 2018.

\bibitem[Carbine et~al.(2018)Carbine, Rodeback, Modersitzki, Miner,
  LeCheminant, and Larson]{carbine2018utility}
K.~A. Carbine, R.~Rodeback, E.~Modersitzki, M.~Miner, J.~D. LeCheminant, and
  M.~J. Larson.
\newblock The utility of event-related potentials (erps) in understanding
  food-related cognition: A systematic review and recommendations.
\newblock \emph{Appetite}, 128:\penalty0 58--78, 2018.

\bibitem[Cepeda-Benito et~al.(2000)Cepeda-Benito, Gleaves, Fern{\'a}ndez, Vila,
  Williams, and Reynoso]{cepeda2000development}
A.~Cepeda-Benito, D.~H. Gleaves, M.~C. Fern{\'a}ndez, J.~Vila, T.~L. Williams,
  and J.~Reynoso.
\newblock The development and validation of spanish versions of the state and
  trait food cravings questionnaires.
\newblock \emph{Behaviour Research and Therapy}, 38\penalty0 (11):\penalty0
  1125--1138, 2000.

\bibitem[Cohen(2013)]{cohen2013statistical}
J.~Cohen.
\newblock \emph{Statistical power analysis for the behavioral sciences}.
\newblock Academic press, 2013.

\bibitem[Cook(1977)]{cook1977detection}
R.~D. Cook.
\newblock Detection of influential observation in linear regression.
\newblock \emph{Technometrics}, 19\penalty0 (1):\penalty0 15--18, 1977.

\bibitem[Das and Nenadic(2009)]{das2009efficient}
K.~Das and Z.~Nenadic.
\newblock An efficient discriminant-based solution for small sample size
  problem.
\newblock \emph{Pattern Recognition}, 42\penalty0 (5):\penalty0 857--866, 2009.

\bibitem[Das et~al.(2009)Das, Rizzuto, and Nenadic]{das2009mental}
K.~Das, D.~S. Rizzuto, and Z.~Nenadic.
\newblock Mental state estimation for brain--computer interfaces.
\newblock \emph{IEEE Transactions on Biomedical Engineering}, 56\penalty0
  (8):\penalty0 2114--2122, 2009.

\bibitem[Das et~al.(2010)Das, Giesbrecht, and Eckstein]{das2010predicting}
K.~Das, B.~Giesbrecht, and M.~P. Eckstein.
\newblock Predicting variations of perceptual performance across individuals
  from neural activity using pattern classifiers.
\newblock \emph{Neuroimage}, 51\penalty0 (4):\penalty0 1425--1437, 2010.

\bibitem[David et~al.(2005)David, Munaf{\`o}, Johansen-Berg, Smith, Rogers,
  Matthews, and Walton]{david2005ventral}
S.~P. David, M.~R. Munaf{\`o}, H.~Johansen-Berg, S.~M. Smith, R.~D. Rogers,
  P.~M. Matthews, and R.~T. Walton.
\newblock Ventral striatum/nucleus accumbens activation to smoking-related
  pictorial cues in smokers and nonsmokers: a functional magnetic resonance
  imaging study.
\newblock \emph{Biological psychiatry}, 58\penalty0 (6):\penalty0 488--494,
  2005.

\bibitem[Delorme and Makeig(2004)]{delorme2004eeglab}
A.~Delorme and S.~Makeig.
\newblock Eeglab: an open source toolbox for analysis of single-trial eeg
  dynamics including independent component analysis.
\newblock \emph{Journal of neuroscience methods}, 134\penalty0 (1):\penalty0
  9--21, 2004.

\bibitem[Di~Chiara and Imperato(1988)]{di1988drugs}
G.~Di~Chiara and A.~Imperato.
\newblock Drugs abused by humans preferentially increase synaptic dopamine
  concentrations in the mesolimbic system of freely moving rats.
\newblock \emph{Proceedings of the National Academy of Sciences}, 85\penalty0
  (14):\penalty0 5274--5278, 1988.

\bibitem[Do et~al.(2011)Do, Wang, King, Abiri, and Nenadic]{do2011brain}
A.~H. Do, P.~T. Wang, C.~E. King, A.~Abiri, and Z.~Nenadic.
\newblock Brain-computer interface controlled functional electrical stimulation
  system for ankle movement.
\newblock \emph{Journal of neuroengineering and rehabilitation}, 8\penalty0
  (1):\penalty0 1--14, 2011.

\bibitem[Do et~al.(2013)Do, Wang, King, Chun, and Nenadic]{do2013brain}
A.~H. Do, P.~T. Wang, C.~E. King, S.~N. Chun, and Z.~Nenadic.
\newblock Brain-computer interface controlled robotic gait orthosis.
\newblock \emph{Journal of neuroengineering and rehabilitation}, 10\penalty0
  (1):\penalty0 1--9, 2013.

\bibitem[Due et~al.(2002)Due, Huettel, Hall, and Rubin]{due2002activation}
D.~L. Due, S.~A. Huettel, W.~G. Hall, and D.~C. Rubin.
\newblock Activation in mesolimbic and visuospatial neural circuits elicited by
  smoking cues: evidence from functional magnetic resonance imaging.
\newblock \emph{American Journal of Psychiatry}, 159\penalty0 (6):\penalty0
  954--960, 2002.

\bibitem[Franken et~al.(2011)Franken, Huijding, Nijs, and van
  Strien]{franken2011electrophysiology}
I.~H. Franken, J.~Huijding, I.~M. Nijs, and J.~W. van Strien.
\newblock Electrophysiology of appetitive taste and appetitive taste
  conditioning in humans.
\newblock \emph{Biological Psychology}, 86\penalty0 (3):\penalty0 273--278,
  2011.

\bibitem[Franklin et~al.(2007)Franklin, Wang, Sciortino, Harper, Li, Ehrman,
  Kampman, O'Brien, Detre, Childress, et~al.]{franklin2007limbic}
T.~R. Franklin, J.~Wang, N.~Sciortino, D.~Harper, Y.~Li, R.~Ehrman, K.~Kampman,
  C.~P. O'Brien, J.~A. Detre, A.~R. Childress, et~al.
\newblock Limbic activation to cigarette smoking cues independent of nicotine
  withdrawal: a perfusion fmri study.
\newblock \emph{Neuropsychopharmacology}, 32\penalty0 (11):\penalty0
  2301--2309, 2007.

\bibitem[Giuliani et~al.(2013)Giuliani, Calcott, and
  Berkman]{giuliani2013piece}
N.~R. Giuliani, R.~D. Calcott, and E.~T. Berkman.
\newblock Piece of cake. cognitive reappraisal of food craving.
\newblock \emph{Appetite}, 64:\penalty0 56--61, 2013.

\bibitem[Giuliani et~al.(2014)Giuliani, Mann, Tomiyama, and
  Berkman]{giuliani2014neural}
N.~R. Giuliani, T.~Mann, A.~J. Tomiyama, and E.~T. Berkman.
\newblock Neural systems underlying the reappraisal of personally craved foods.
\newblock \emph{Journal of cognitive neuroscience}, 26\penalty0 (7):\penalty0
  1390--1402, 2014.

\bibitem[Haber and Knutson(2010)]{haber2010reward}
S.~N. Haber and B.~Knutson.
\newblock The reward circuit: linking primate anatomy and human imaging.
\newblock \emph{Neuropsychopharmacology}, 35\penalty0 (1):\penalty0 4--26,
  2010.

\bibitem[Hachl et~al.(2003)Hachl, Hempel, and Pietrowsky]{hachl2003erps}
P.~Hachl, C.~Hempel, and R.~Pietrowsky.
\newblock Erps to stimulus identification in persons with restrained eating
  behavior.
\newblock \emph{International Journal of Psychophysiology}, 49\penalty0
  (2):\penalty0 111--121, 2003.

\bibitem[Hajcak et~al.(2010)Hajcak, MacNamara, and Olvet]{hajcak2010event}
G.~Hajcak, A.~MacNamara, and D.~M. Olvet.
\newblock Event-related potentials, emotion, and emotion regulation: an
  integrative review.
\newblock \emph{Developmental neuropsychology}, 35\penalty0 (2):\penalty0
  129--155, 2010.

\bibitem[Harris et~al.(2013)Harris, Hare, and Rangel]{harris2013temporally}
A.~Harris, T.~Hare, and A.~Rangel.
\newblock Temporally dissociable mechanisms of self-control: early attentional
  filtering versus late value modulation.
\newblock \emph{Journal of Neuroscience}, 33\penalty0 (48):\penalty0
  18917--18931, 2013.

\bibitem[Haynes and Rees(2005)]{haynes2005predicting}
J.-D. Haynes and G.~Rees.
\newblock Predicting the orientation of invisible stimuli from activity in
  human primary visual cortex.
\newblock \emph{Nature neuroscience}, 8\penalty0 (5):\penalty0 686--691, 2005.

\bibitem[Jansen et~al.(2010)Jansen, Stegerman, Roefs, Nederkoorn, and
  Havermans]{jansen2010decreased}
A.~Jansen, S.~Stegerman, A.~Roefs, C.~Nederkoorn, and R.~Havermans.
\newblock Decreased salivation to food cues in formerly obese successful
  dieters.
\newblock \emph{Psychotherapy and psychosomatics}, 79\penalty0 (4):\penalty0
  257, 2010.

\bibitem[Jansen et~al.(2016)Jansen, Schyns, Bongers, and van~den
  Akker]{jansen2016lab}
A.~Jansen, G.~Schyns, P.~Bongers, and K.~van~den Akker.
\newblock From lab to clinic: extinction of cued cravings to reduce overeating.
\newblock \emph{Physiology \& Behavior}, 162:\penalty0 174--180, 2016.

\bibitem[Kamarajan et~al.(2010)Kamarajan, Rangaswamy, Tang, Chorlian, Pandey,
  Roopesh, Manz, Saunders, Stimus, and Porjesz]{kamarajan2010dysfunctional}
C.~Kamarajan, M.~Rangaswamy, Y.~Tang, D.~B. Chorlian, A.~K. Pandey, B.~N.
  Roopesh, N.~Manz, R.~Saunders, A.~T. Stimus, and B.~Porjesz.
\newblock Dysfunctional reward processing in male alcoholics: an erp study
  during a gambling task.
\newblock \emph{Journal of psychiatric research}, 44\penalty0 (9):\penalty0
  576--590, 2010.

\bibitem[Kelley and Preacher(2012)]{kelley2012effect}
K.~Kelley and K.~J. Preacher.
\newblock On effect size.
\newblock \emph{Psychological methods}, 17\penalty0 (2):\penalty0 137, 2012.

\bibitem[Killick and Eckley(2014)]{killick2014changepoint}
R.~Killick and I.~Eckley.
\newblock changepoint: An r package for changepoint analysis.
\newblock \emph{Journal of statistical software}, 58\penalty0 (3):\penalty0
  1--19, 2014.

\bibitem[Killick et~al.(2012)Killick, Fearnhead, and
  Eckley]{killick2012optimal}
R.~Killick, P.~Fearnhead, and I.~A. Eckley.
\newblock Optimal detection of changepoints with a linear computational cost.
\newblock \emph{Journal of the American Statistical Association}, 107\penalty0
  (500):\penalty0 1590--1598, 2012.

\bibitem[Kober et~al.(2010{\natexlab{a}})Kober, Kross, Mischel, Hart, and
  Ochsner]{kober2010regulation}
H.~Kober, E.~F. Kross, W.~Mischel, C.~L. Hart, and K.~N. Ochsner.
\newblock Regulation of craving by cognitive strategies in cigarette smokers.
\newblock \emph{Drug and alcohol dependence}, 106\penalty0 (1):\penalty0
  52--55, 2010{\natexlab{a}}.

\bibitem[Kober et~al.(2010{\natexlab{b}})Kober, Mende-Siedlecki, Kross, Weber,
  Mischel, Hart, and Ochsner]{kober2010prefrontal}
H.~Kober, P.~Mende-Siedlecki, E.~F. Kross, J.~Weber, W.~Mischel, C.~L. Hart,
  and K.~N. Ochsner.
\newblock Prefrontal--striatal pathway underlies cognitive regulation of
  craving.
\newblock \emph{Proceedings of the National Academy of Sciences}, 107\penalty0
  (33):\penalty0 14811--14816, 2010{\natexlab{b}}.

\bibitem[Kong et~al.(2015)Kong, Zhang, and Chen]{kong2015inhibition}
F.~Kong, Y.~Zhang, and H.~Chen.
\newblock Inhibition ability of food cues between successful and unsuccessful
  restrained eaters: a two-choice oddball task.
\newblock \emph{PLoS One}, 10\penalty0 (4):\penalty0 e0120522, 2015.

\bibitem[Konova et~al.(2018)Konova, Louie, and
  Glimcher]{konova2018computational}
A.~B. Konova, K.~Louie, and P.~W. Glimcher.
\newblock The computational form of craving is a selective multiplication of
  economic value.
\newblock \emph{Proceedings of the National Academy of Sciences}, 115\penalty0
  (16):\penalty0 4122--4127, 2018.

\bibitem[Lafay et~al.(2001)Lafay, Thomas, Mennen, Charles, Eschwege, and
  Borys]{lafay2001gender}
L.~Lafay, F.~Thomas, L.~Mennen, M.~A. Charles, E.~Eschwege, and J.-M. Borys.
\newblock Gender differences in the relation between food cravings and mood in
  an adult community: Results from the fleurbaix laventie ville sante study.
\newblock \emph{International Journal of Eating Disorders}, 29\penalty0
  (2):\penalty0 195--204, 2001.

\bibitem[Lapenta et~al.(2014)Lapenta, Di~Sierve, de~Macedo, Fregni, and
  Boggio]{lapenta2014transcranial}
O.~M. Lapenta, K.~Di~Sierve, E.~C. de~Macedo, F.~Fregni, and P.~S. Boggio.
\newblock Transcranial direct current stimulation modulates erp-indexed
  inhibitory control and reduces food consumption.
\newblock \emph{Appetite}, 83:\penalty0 42--48, 2014.

\bibitem[Ledoux et~al.(2013)Ledoux, Nguyen, Bakos-Block, and
  Bordnick]{ledoux2013using}
T.~Ledoux, A.~S. Nguyen, C.~Bakos-Block, and P.~Bordnick.
\newblock Using virtual reality to study food cravings.
\newblock \emph{Appetite}, 71:\penalty0 396--402, 2013.

\bibitem[Liu et~al.(2014)Liu, Wisdom, Roberto, Liu, and Ubel]{liu2014using}
P.~J. Liu, J.~Wisdom, C.~A. Roberto, L.~J. Liu, and P.~A. Ubel.
\newblock Using behavioral economics to design more effective food policies to
  address obesity.
\newblock \emph{Applied Economic Perspectives and Policy}, 36\penalty0
  (1):\penalty0 6--24, 2014.

\bibitem[Lowe et~al.(2018)Lowe, Staines, Manocchio, and
  Hall]{lowe2018neurocognitive}
C.~J. Lowe, W.~R. Staines, F.~Manocchio, and P.~A. Hall.
\newblock The neurocognitive mechanisms underlying food cravings and snack food
  consumption. a combined continuous theta burst stimulation (ctbs) and eeg
  study.
\newblock \emph{Neuroimage}, 177:\penalty0 45--58, 2018.

\bibitem[Luijten et~al.(2014)Luijten, Machielsen, Veltman, Hester, de~Haan, and
  Franken]{luijten2014systematic}
M.~Luijten, M.~W. Machielsen, D.~J. Veltman, R.~Hester, L.~de~Haan, and I.~H.
  Franken.
\newblock Systematic review of erp and fmri studies investigating inhibitory
  control and error processing in people with substance dependence and
  behavioural addictions.
\newblock \emph{Journal of psychiatry \& neuroscience}, 2014.

\bibitem[Meule et~al.(2013)Meule, K{\"u}bler, and Blechert]{meule2013time}
A.~Meule, A.~K{\"u}bler, and J.~Blechert.
\newblock Time course of electrocortical food-cue responses during cognitive
  regulation of craving.
\newblock \emph{Frontiers in Psychology}, 4:\penalty0 669, 2013.

\bibitem[Mischel and Baker(1975)]{mischel1975cognitive}
W.~Mischel and N.~Baker.
\newblock Cognitive appraisals and transformations in delay behavior.
\newblock \emph{Journal of personality and social psychology}, 31\penalty0
  (2):\penalty0 254, 1975.

\bibitem[Nijs et~al.(2007)Nijs, Franken, and Muris]{nijs2007modified}
I.~M. Nijs, I.~H. Franken, and P.~Muris.
\newblock The modified trait and state food-cravings questionnaires:
  development and validation of a general index of food craving.
\newblock \emph{Appetite}, 49\penalty0 (1):\penalty0 38--46, 2007.

\bibitem[Nijs et~al.(2008)Nijs, Franken, and Muris]{nijs2008food}
I.~M. Nijs, I.~H. Franken, and P.~Muris.
\newblock Food cue-elicited brain potentials in obese and healthy-weight
  individuals.
\newblock \emph{Eating behaviors}, 9\penalty0 (4):\penalty0 462--470, 2008.

\bibitem[O'Doherty et~al.(2006)O'Doherty, Buchanan, Seymour, and
  Dolan]{o2006predictive}
J.~P. O'Doherty, T.~W. Buchanan, B.~Seymour, and R.~J. Dolan.
\newblock Predictive neural coding of reward preference involves dissociable
  responses in human ventral midbrain and ventral striatum.
\newblock \emph{Neuron}, 49\penalty0 (1):\penalty0 157--166, 2006.

\bibitem[Patrick et~al.(2006)Patrick, Bernat, Malone, Iacono, Krueger, and
  McGue]{patrick2006p300}
C.~J. Patrick, E.~M. Bernat, S.~M. Malone, W.~G. Iacono, R.~F. Krueger, and
  M.~McGue.
\newblock P300 amplitude as an indicator of externalizing in adolescent males.
\newblock \emph{Psychophysiology}, 43\penalty0 (1):\penalty0 84--92, 2006.

\bibitem[Pelchat et~al.(2004)Pelchat, Johnson, Chan, Valdez, and
  Ragland]{pelchat2004images}
M.~L. Pelchat, A.~Johnson, R.~Chan, J.~Valdez, and J.~D. Ragland.
\newblock Images of desire: food-craving activation during fmri.
\newblock \emph{Neuroimage}, 23\penalty0 (4):\penalty0 1486--1493, 2004.

\bibitem[Philiastides et~al.(2006)Philiastides, Ratcliff, and
  Sajda]{philiastides2006neural}
M.~G. Philiastides, R.~Ratcliff, and P.~Sajda.
\newblock Neural representation of task difficulty and decision making during
  perceptual categorization: a timing diagram.
\newblock \emph{Journal of Neuroscience}, 26\penalty0 (35):\penalty0
  8965--8975, 2006.

\bibitem[Roy et~al.(2021)Roy, Mazumder, and Das]{roy2021wisdom}
T.~S. Roy, S.~Mazumder, and K.~Das.
\newblock Wisdom of crowds benefits perceptual decision making across
  difficulty levels.
\newblock \emph{Scientific reports}, 11\penalty0 (1):\penalty0 1--13, 2021.

\bibitem[Saha~Roy et~al.(2020)Saha~Roy, Giri, Saha~Chowdhury, Mazumder, and
  Das]{saha2020our}
T.~Saha~Roy, B.~Giri, A.~Saha~Chowdhury, S.~Mazumder, and K.~Das.
\newblock How our perception and confidence are altered using decision cues.
\newblock \emph{Frontiers in neuroscience}, 13:\penalty0 1371, 2020.

\bibitem[Sarlo et~al.(2013)Sarlo, {\"U}bel, Leutgeb, and
  Schienle]{sarlo2013cognitive}
M.~Sarlo, S.~{\"U}bel, V.~Leutgeb, and A.~Schienle.
\newblock Cognitive reappraisal fails when attempting to reduce the appetitive
  value of food: An erp study.
\newblock \emph{Biological psychology}, 94\penalty0 (3):\penalty0 507--512,
  2013.

\bibitem[Schienle et~al.(2009)Schienle, Sch{\"a}fer, Hermann, and
  Vaitl]{schienle2009binge}
A.~Schienle, A.~Sch{\"a}fer, A.~Hermann, and D.~Vaitl.
\newblock Binge-eating disorder: reward sensitivity and brain activation to
  images of food.
\newblock \emph{Biological psychiatry}, 65\penalty0 (8):\penalty0 654--661,
  2009.

\bibitem[Schwab et~al.(2017)Schwab, Giraldo, Spiegl, and
  Schienle]{schwab2017disgust}
D.~Schwab, M.~Giraldo, B.~Spiegl, and A.~Schienle.
\newblock Disgust evoked by strong wormwood bitterness influences the
  processing of visual food cues in women: An erp study.
\newblock \emph{Appetite}, 108:\penalty0 51--56, 2017.

\bibitem[Stockburger et~al.(2009)Stockburger, Renner, Weike, Hamm, and
  Schupp]{stockburger2009vegetarianism}
J.~Stockburger, B.~Renner, A.~I. Weike, A.~O. Hamm, and H.~T. Schupp.
\newblock Vegetarianism and food perception. selective visual attention to meat
  pictures.
\newblock \emph{Appetite}, 52\penalty0 (2):\penalty0 513--516, 2009.

\bibitem[Svaldi et~al.(2015)Svaldi, Tuschen-Caffier, Biehl, Gschwendtner, Wolz,
  and Naumann]{svaldi2015effects}
J.~Svaldi, B.~Tuschen-Caffier, S.~C. Biehl, K.~Gschwendtner, I.~Wolz, and
  E.~Naumann.
\newblock Effects of two cognitive regulation strategies on the processing of
  food cues in high restrained eaters. an event-related potential study.
\newblock \emph{Appetite}, 92:\penalty0 269--277, 2015.

\bibitem[Tang et~al.(2012)Tang, Fellows, Small, and Dagher]{tang2012food}
D.~Tang, L.~Fellows, D.~Small, and A.~Dagher.
\newblock Food and drug cues activate similar brain regions: a meta-analysis of
  functional mri studies.
\newblock \emph{Physiology \& behavior}, 106\penalty0 (3):\penalty0 317--324,
  2012.

\bibitem[Versace et~al.(2016)Versace, Kypriotakis, Basen-Engquist, and
  Schembre]{versace2016heterogeneity}
F.~Versace, G.~Kypriotakis, K.~Basen-Engquist, and S.~M. Schembre.
\newblock Heterogeneity in brain reactivity to pleasant and food cues: evidence
  of sign-tracking in humans.
\newblock \emph{Social Cognitive and Affective Neuroscience}, 11\penalty0
  (4):\penalty0 604--611, 2016.

\bibitem[Waters et~al.(2001)Waters, Hill, and Waller]{waters2001internal}
A.~Waters, A.~Hill, and G.~Waller.
\newblock Internal and external antecedents of binge eating episodes in a group
  of women with bulimia nervosa.
\newblock \emph{International Journal of Eating Disorders}, 29\penalty0
  (1):\penalty0 17--22, 2001.

\bibitem[Wessel and Aron(2015)]{wessel2015s}
J.~R. Wessel and A.~R. Aron.
\newblock It's not too late: The onset of the frontocentral p 3 indexes
  successful response inhibition in the stop-signal paradigm.
\newblock \emph{Psychophysiology}, 52\penalty0 (4):\penalty0 472--480, 2015.

\bibitem[Wilson et~al.(2016)Wilson, Buckley, Buckley, and
  Bogomolova]{wilson2016nudging}
A.~L. Wilson, E.~Buckley, J.~D. Buckley, and S.~Bogomolova.
\newblock Nudging healthier food and beverage choices through salience and
  priming. evidence from a systematic review.
\newblock \emph{Food Quality and Preference}, 51:\penalty0 47--64, 2016.

\bibitem[Wilson et~al.(2007)Wilson, Grilo, and
  Vitousek]{wilson2007psychological}
G.~T. Wilson, C.~M. Grilo, and K.~M. Vitousek.
\newblock Psychological treatment of eating disorders.
\newblock \emph{American Psychologist}, 62\penalty0 (3):\penalty0 199, 2007.

\bibitem[Wilson et~al.(2005)Wilson, Sayette, Delgado, and
  Fiez]{wilson2005instructed}
S.~J. Wilson, M.~A. Sayette, M.~R. Delgado, and J.~A. Fiez.
\newblock Instructed smoking expectancy modulates cue-elicited neural activity:
  a preliminary study.
\newblock \emph{Nicotine \& tobacco research}, 7\penalty0 (4):\penalty0
  637--645, 2005.

\bibitem[Wolz et~al.(2017)Wolz, Sauvaget, Granero, Mestre-Bach, Ba{\~n}o,
  Mart{\'\i}n-Romera, de~Las~Heras, Jim{\'e}nez-Murcia, Jansen, Roefs,
  et~al.]{wolz2017subjective}
I.~Wolz, A.~Sauvaget, R.~Granero, G.~Mestre-Bach, M.~Ba{\~n}o,
  V.~Mart{\'\i}n-Romera, M.~V. de~Las~Heras, S.~Jim{\'e}nez-Murcia, A.~Jansen,
  A.~Roefs, et~al.
\newblock Subjective craving and event-related brain response to olfactory and
  visual chocolate cues in binge-eating and healthy individuals.
\newblock \emph{Scientific reports}, 7\penalty0 (1):\penalty0 1--10, 2017.

\bibitem[Zorjan et~al.(2021)Zorjan, Gremsl, and Schienle]{zorjan2021changing}
S.~Zorjan, A.~Gremsl, and A.~Schienle.
\newblock Changing the visualization of food to reduce food cue reactivity: An
  event-related potential study.
\newblock \emph{Biological Psychology}, 164:\penalty0 108173, 2021.

\end{thebibliography}
	
\end{document}